\documentclass[aps,superscriptaddress,twocolumn]{revtex4-2}
\usepackage[utf8]{inputenc}
\usepackage{amsmath,amssymb}
\usepackage{graphicx}
\usepackage{xcolor, soul}
\usepackage{hyperref}
\usepackage[capitalise]{cleveref}
\usepackage{todonotes}
\usepackage{braket}

\begin{document}
\title{A mechanical quantum memory for microwave photons}
\date{\today}
\author{Alkım B. Bozkurt}
\thanks{These authors contributed equally.}
\affiliation{The Gordon and Betty Moore Laboratory of Engineering, California Institute of Technology, Pasadena, California 91125}
\affiliation{Institute for Quantum Information and Matter, California Institute of Technology, Pasadena, California 91125}
\author{Omid Golami}
\thanks{These authors contributed equally.}
\affiliation{The Gordon and Betty Moore Laboratory of Engineering, California Institute of Technology, Pasadena, California 91125}
\affiliation{Institute for Quantum Information and Matter, California Institute of Technology, Pasadena, California 91125}

\author{Yue Yu}
\affiliation{The Gordon and Betty Moore Laboratory of Engineering, California Institute of Technology, Pasadena, California 91125}

\author{Hao Tian}
\affiliation{The Gordon and Betty Moore Laboratory of Engineering, California Institute of Technology, Pasadena, California 91125}
\affiliation{Institute for Quantum Information and Matter, California Institute of Technology, Pasadena, California 91125}

\author{Mohammad Mirhosseini}
\email{mohmir@caltech.edu}
\homepage{http://qubit.caltech.edu}
\affiliation{The Gordon and Betty Moore Laboratory of Engineering, California Institute of Technology, Pasadena, California 91125}
\affiliation{Institute for Quantum Information and Matter, California Institute of Technology, Pasadena, California 91125}
\begin{abstract}

Long-lived mechanical oscillators are actively pursued as critical resources for quantum storage, sensing, and transduction. However, achieving deterministic quantum control while limiting mechanical dissipation remains a persistent challenge. Here, we demonstrate strong coupling between a transmon superconducting qubit and an ultra-long-lived nanomechanical oscillator ($T_\text{1} \approx 25 \text{ ms}$ at 5 GHz, $Q \approx 0.8 \times 10^9$) by leveraging the low acoustic loss in silicon and phononic bandgap engineering. The qubit-oscillator system achieves large cooperativity ($C_{T_1}\approx 1.5\times10^5$, $C_{T_2}\approx 150$), enabling the generation of non-classical states and the investigation of mechanisms underlying mechanical decoherence. We show that dynamical decoupling—implemented through the qubit—can mitigate decoherence, leading to a mechanical coherence time of $T_2\approx 1 \text{ ms}$. These findings extend the exceptional storage capabilities of mechanical oscillators to the quantum regime, putting them forward as compact bosonic elements for future applications in quantum computing and metrology.

\end{abstract}

\maketitle

Harmonic motion, originating from elastic forces in solid matter, plays a central role in classical science and technology: mechanical oscillators are widely utilized in timekeeping \cite{nguyenMEMSTechnologyTiming2007}, telecommunication \cite{ruppel2017}, navigation \cite{syedNewMultipositionCalibration2007}, and sensing \cite{osullivan1999}. In the past decade, the advent of mechanical systems in the quantum regime has opened up a new frontier with potentially transformative impact in quantum-enhanced sensing \cite{mason2019,mccormick2019}, fundamental tests of quantum gravity \cite{pikovski2012,arndt2014}, quantum computing \cite{qiao2023, yang2024}, and transduction of quantum signals \cite{barzanjehOptomechanicsQuantumTechnologies2022}.

A critical physical property for harnessing mechanical oscillators in both classical and quantum systems is their dissipation, often quantified by the quality factor, $Q$. Extensive work in understanding the physical origins of acoustic loss and advances in engineering have recently culminated in mechanical oscillators with exceptionally high quality factors, in excess of one billion  \cite{galliou2013,engelsen2024,maccabe2020}. Leveraging these systems in the quantum regime, however, requires either nonlinear interactions or coupling to a quantum two-level-system, \emph{i.e.}, a qubit. Piezoelectric coupling to superconducting qubits has been particularly fruitful, demonstrating deterministic preparation, manipulation, and measurement of non-classical states of motion \cite{satzinger2018,bienfait2020,chu2018,bild2023, wollack2022a}. Despite the promises, the challenges in integrating piezoelectric materials with qubits have so far limited the mechanical quality factors to a few million \cite{yang2024}, precluding quantum control of ultra-high-$Q$ oscillators.

In this work, we present an alternative mechanism for coupling superconducting qubits to mechanical oscillators. Our approach achieves electromechanical interaction through the nonlinearity of the electrostatic force in nanoscale devices. Crucially, this material-agnostic approach enables the coupling of qubits to mechanical oscillators exclusively made of single-crystal silicon. Relying on exceptionally low material loss in silicon \cite{mcguigan1978}, combined with the suppression of extrinsic energy loss using phononic band gaps, we demonstrate an intrinsic mechanical quality factor of 0.8 billion, measured at the single-phonon level ($T_1 = 25 \pm 2 \text{ ms}$ at 4.9 GHz). We demonstrate that the qubit-oscillator system is in the strong-coupling regime, allowing for deterministic quantum operations. This unprecedented access to qubit control in an ultra-high-$Q$ oscillator, enables us to directly probe the microscopic origins of mechanical decoherence often attributed to two-level-system (TLS) defects, where we observe distinctive signatures of individual defects.  Finally, we show that mechanical dephasing can be effectively mitigated with dynamical decoupling, demonstrating mechanical coherence times reaching the millisecond range ($T_2^\text{CP2} = 1.02\pm 0.15 \:\text{ms}$).

Importantly, our measured mechanical lifetimes surpass those of superconducting qubits and planar microwave resonators by over an order of magnitude \cite{ganjam2024,crowley2023,tuokkolaMethodsAchieveMillisecond2024}, while our oscillators occupy footprints several orders of magnitude smaller. These results put forth mechanical oscillators as long-lived bosonic elements for storing microwave-frequency quantum states for future applications in quantum computing and metrology \cite{hann2019, chamberland2022, cai2021,deng2024}.

\begin{figure*}[ht]
    \centering
    \includegraphics[width=\textwidth]{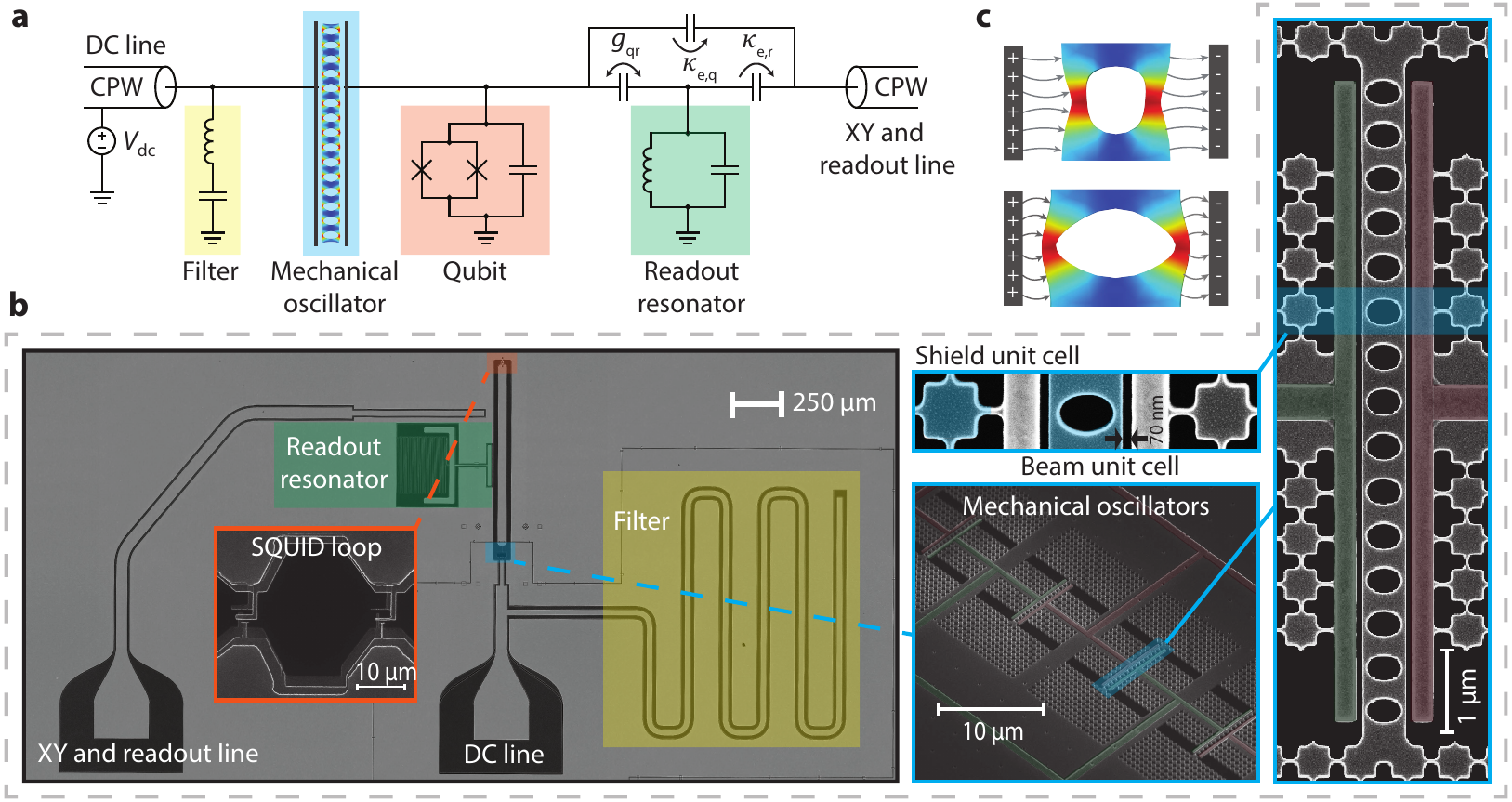}
    \caption{\textbf{Experimental Setup.} \textbf{a,} Circuit diagram of the coupled qubit-oscillator system. \textbf{b,} False colored optical microscope image of the device, showing the SQUID loop of the transmon qubit (orange), readout resonator (green), filter (yellow), and mechanical oscillators (blue) alongside the coplanar waveguides (CPW) that deliver DC and RF signals. Insets: False-colored tilted SEM image of multiple mechanical oscillators, with the red (green) shaded electrodes attached to the qubit (DC line). SEM image of a single mechanical oscillator, with the blue shaded section showing a single phononic crystal unit cell, phononic shields, and the vacuum gap between the oscillator and the electrodes. \textbf{c,} Simulated displacement profile of a unit cell of the phononic crystal resonator, alongside an illustration of electric field lines between the electrodes and the nanobeam. The two images demonstrate the `breathing` motion separated in time by half an oscillation period.}
    \label{fig:setup} 
\end{figure*}

\section*{System Description}

Our system consists of a transmon superconducting qubit that is galvanically connected to a mechanically compliant vacuum gap capacitor (see \cref{fig:setup}a).  In this setup, a mechanical resonance can be formed within the capacitor by engineering the motional boundary conditions \cite{bozkurt2023}. Applying an electrostatic field to the moving capacitor with an external source gives rise to an electrical dipole that oscillates with mechanical motion. This oscillating dipole interacts with the electric field from the qubit \cite{pirkkalainen2013,ma2021,lahaye2009}, giving rise to a coupled system described by the Hamiltonian  

\begin{equation}
    \hat{H}/\hbar = \frac{\omega_\text{q}}{2}\hat{\sigma}_\text{z}  + \omega_\text{m}\hat{b}^{\dagger}\hat{b} + g_\text{em}\left(\hat{\sigma}_+\hat{b}+\hat{\sigma}_-\hat{b}^{\dagger}\right).
\end{equation}
Here, $\omega_\text{q}$ ($\omega_\text{m}$) is the transition (resonance) frequency of the qubit (oscillator), $\sigma_\text{-}$ ($\hat{b}$) is the qubit (oscillator) annihilation operator, and $g_\text{em}$ is the electromechanical coupling rate, which depends on the external voltage bias ($g_\text{em} = g_0V_\text{dc}$, see \cref{app:sup_hamiltonian}). When the resonance condition is satisfied (i.e, $\omega_\text{q} = \omega_\text{m}$), the electromechanical interaction enables the exchange of phonons and qubit excitations (i.e., microwave photons). 

We implement this system on a silicon-on-insulator platform, where a planar transmon qubit is fabricated on a suspended membrane accommodating moving parts, as shown in \cref{fig:setup}b (for details, see \cref{app:sup_fab}).  The moving capacitor is created by a pair of electrodes surrounding a phononic crystal resonator.  We use finite-element modeling to optimize the geometry of the phononic crystal for maximum electromechanical interaction (see \cref{si_mech_design}). \Cref{fig:setup}c shows the mechanical mode profile, where in-plane `breathing' motion leads to the variations of the vacuum gap along the beam. Crucially, our capacitor design enables the detection of mechanical motion through near-field components of the electric field while avoiding mechanical energy within the metal electrodes (which have previously been identified as sources of considerable mechanical loss \cite{wollack2021b,mason1947}). The mechanical oscillator is also surrounded by phononic shields with a full band gap in the vicinity of the mechanical resonance to avoid clamping loss (\cite{maccabe2020}, see \cref{si_mech_design}). The compact design of the nanomechanical oscillators allows us to connect an array of them to a single transmon qubit. We utilize flux tuning by an off-chip magnet to precisely adjust the qubit frequency to resonance with only one mechanical oscillator while providing sufficiently large detunings to avoid interactions with the other oscillators in the array.  We apply the electrostatic field to the moving capacitor via a coplanar waveguide. However, in this geometry, the qubit experiences radiative decay to the coplanar waveguide via the motional capacitance. To mitigate this parasitic effect, we integrate a notch-type microwave filter on the same chip engineered to reflect signals at the qubit frequency while remaining transparent to zero-frequency biasing fields (see \cref{si_mw_design}). The qubit is also coupled to a microwave resonator for dispersive readout.

\begin{figure}[t]
    \centering
    \includegraphics[width=0.5\textwidth]{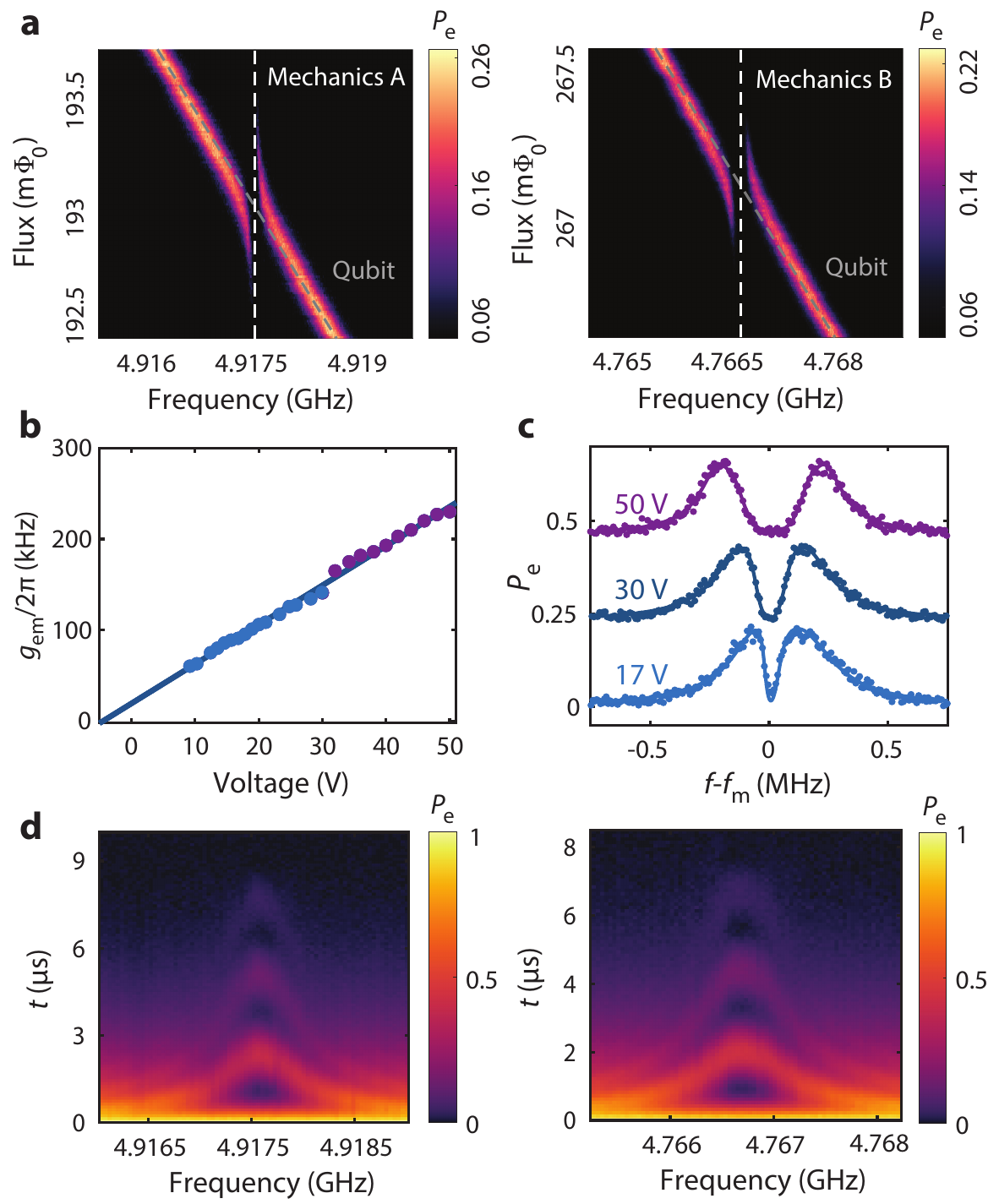}
    \caption{\textbf{Coherent qubit-oscillator interaction.} \textbf{a,} Microwave spectroscopy of the qubit under flux tuning, exhibiting avoided crossings with oscillators A and B. Dashed lines indicate bare qubit and oscillator frequencies. \textbf{b,} The electromechanical interaction rate of device B as a function of the applied DC voltage, along with a linear theory fit. The blue (purple) data points are extracted from the spectroscopy (time domain) traces. \textbf{c,} Qubit resonant spectroscopy with mechanics B at different voltages, showing the transition from weak to strong coupling regime (with the measurements at 30 V and 50 V offset vertically for visual clarity). A fit to the measurement at 50 V provides the maximum interaction rates, $g_\text{em,A}/2\pi = 200 \text{ kHz}$ (data not shown) and $g_\text{em,B}/2\pi = 230 \text{ kHz}$.  \textbf{d,} Qubit time dynamics exhibiting vacuum Rabi oscillations between the qubit and the oscillator. $P_\text{e}$ is the qubit excited state population and $\Phi_0$ is the flux quantum. The data presented in \textbf{a} and \textbf{d} is obtained at a voltage bias of 50 V.}
    \label{fig:qm_interaction}
\end{figure}

\begin{figure}[t]
    \centering
    \includegraphics[width=0.5\textwidth]{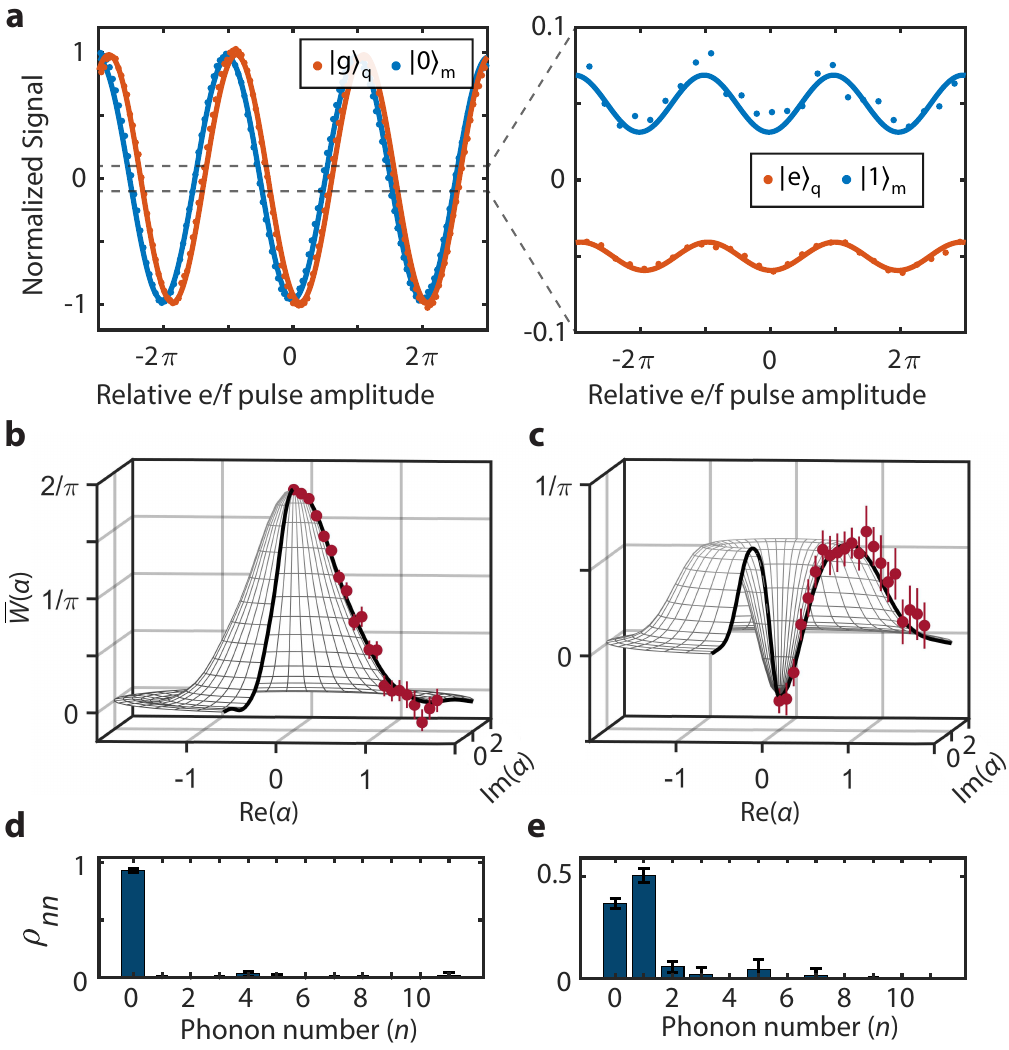}
    \caption{\textbf{Quantum state preparation and tomography.} \textbf{a,} The thermal population of the qubit's $\ket{g}_\text{q}$ ($\ket{e}_\text{q}$) level is measured through the amplitude of driven Rabi oscillations in the e-f manifold, as shown in the left (right) panel. For measuring the population in $\ket{g}_\text{q}$, we initiate the experiment with a g-e $\pi$-pulse (see \cref{si_thermometry}). The thermal population of the oscillator's $\ket{0}_\text{m}$ ($\ket{1}_\text{m}$) energy level is extracted from a similar experiment conducted via the qubit after performing a mechanics-qubit state swap, as shown in the left (right) panel. The data on the left (right) panel is offset horizontally (vertically) for better visibility. \textbf{b,} Wigner tomogram of the mechanical oscillator prepared in the ground state and \textbf{c,} the single-phonon excited state, showing $\overline{W}(0) = -0.10\pm0.02$. Red data points are calculated by fitting the experimental data with 2$\sigma$ error bars; the 3-D grids denote the reconstructed Wigner distribution function. \textbf{d} and \textbf{e,} Diagonal elements of the reconstructed density matrix for the ground and excited mechanical states, respectively. All the data is taken at 50 V using mechanics B.}
    \label{fig:Wig}
\end{figure}

\begin{figure*}[ht]
    \centering
    \includegraphics[width=\textwidth]{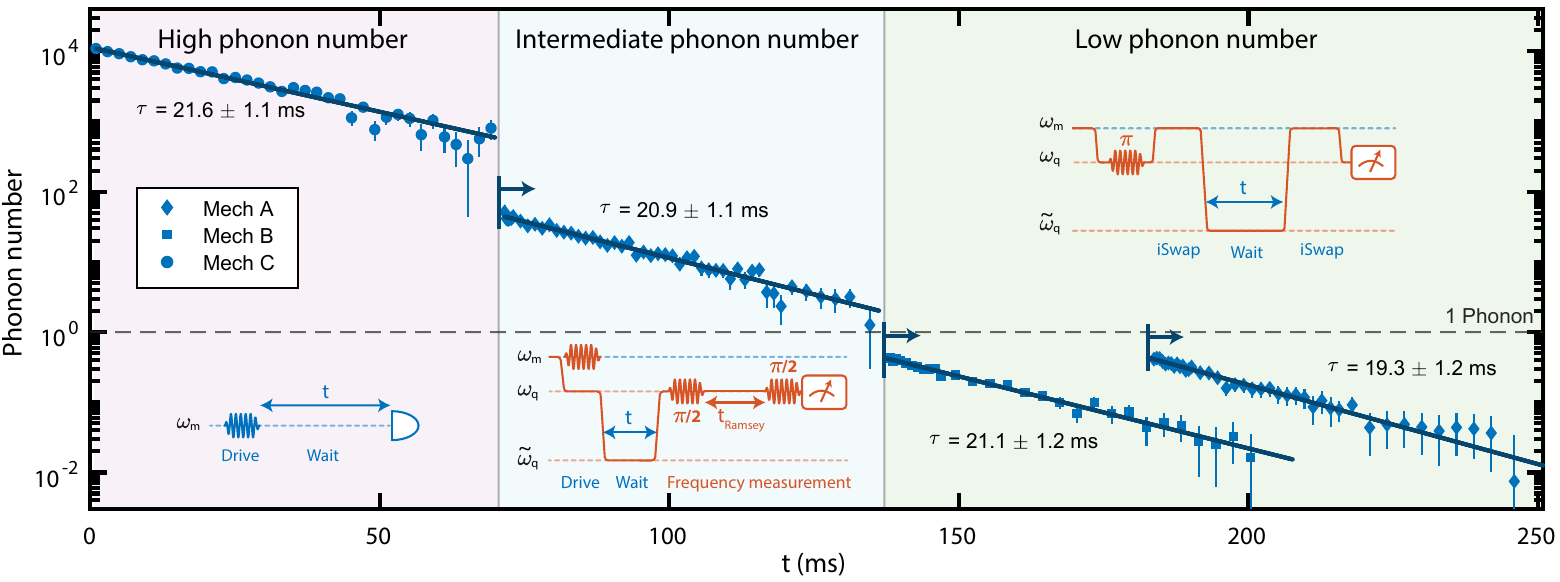}
    \caption{\textbf{Mechanical lifetime.} We employ three distinct measurement methods, highlighted by shaded areas, to capture decay time measurements across various phonon occupancy levels, with each trace offset in time for visual clarity. The lifetimes obtained by exponential fits of the form $ae^{-t/\tau}$ are displayed for each measurement and error bars indicate $1\sigma$ confidence interval.  For experiments in the range below 1 phonon, see the main text. For the experiment in the approximate range of 1-100 phonons (mechanics A), we employ the qubit as a phonon-number detector by operation in the dispersive regime \cite{jeffreyFastAccurateState2014, sankSystemCharacterizationDispersive2024}. At the highest phonon number range, we measure an array of identical mechanical oscillators on the same chip (represented by mechanics C) with electromechanical readout through a directly connected microwave waveguide (see \cref{si_high_PhN}). All measurements are conducted at a bias voltage of 40 V. Insets show the pulse sequences used for each measurement.
    For qubit-based measurements, we use a microwave tone for shifting the qubit's transition frequency away from the oscillator (via the ac Stark effect) to minimize the inverse Purcell decay \cite{zhangEngineeringBilinearMode2019}. For mech A, we characterize and subtract the small contribution from the inverse Purcell decay to find the intrinsic lifetime ($25\pm2$ ms, see \cref{si_Purcell}).}
    \label{fig:lifetime} 
\end{figure*}

\section*{Qubit-Oscillator Interactions}

We characterize the fabricated devices in a dilution fridge with a base temperature of 10 mK. With a voltage applied to the motional capacitance, we conduct spectroscopy as we slowly tune the qubit frequency. As shown in \cref{fig:qm_interaction}a, this measurement reveals two avoided mode crossings, from which we extract the resonance frequencies of two mechanical oscillators ($\omega_\text{m,A}/2\pi = 4.9176 \text{ GHz}$ and $\omega_\text{m,B}/2\pi =4.7667 \text{ GHz}$) and the electromechanical interaction rates. The interaction rates scale linearly with the voltage bias (see \cref{fig:qm_interaction}b) in line with our device modeling (see \cref{si_mech_design}). Crucially, the interaction can be enhanced sufficiently to yield a pair of resolved hybridized modes at resonance (see \cref{fig:qm_interaction}c), marking the entry to the strong-coupling regime. We confirm operation in this regime by performing time domain measurements, where we observe vacuum Rabi oscillations (see \cref{fig:qm_interaction}d). These oscillations demonstrate the coherent exchange of photons and phonons at the single quantum level, which is the basis for deterministic quantum control of the mechanical oscillator via the qubit.

Beyond strong interactions, quantum operations often require preparing mechanical oscillators in the quantum ground state. Although the cryostat temperature in our experiment is sufficiently low for this purpose ($\hbar\omega_\text{m}\gg k_\mathrm{B}T_\mathrm{fridge}$), the strong electrostatic field at the device can potentially lead to an increased effective temperature via parasitic effects \cite{rouxinol2016a,catto2022}. To investigate these potential effects, we quantify the residual thermal populations of the qubit and mechanical resonator in our system using established thermometry methods \cite{geerlingsDemonstratingDrivenReset2013} 
(see \cref{si_thermometry}).  These measurements (see \cref{fig:Wig}a) yield small and comparable mode temperatures for the qubit ($60 \pm 2$ mK) and the mechanical oscillator ($72 \pm 9$ mK). Importantly, we observe no significant change in the qubit temperature in the absence of the bias voltage ($62\pm1$ mK), effectively ruling out heating from the electrostatic biasing field.

The combination of strong qubit-oscillator interaction and ground-state operation in our system enables the preparation and measurement of non-classical motional states \cite{satzinger2018,chu2018}. For state preparation, we initially prepare the qubit in the desired quantum state and then transfer this state to the mechanical oscillator. Subsequently, we perform Wigner tomography by displacing the mechanical quantum state using a resonant drive and determining its parity through fits of qubit-mechanics Rabi oscillations (see \cref{si_Wigner}). \Cref{fig:Wig}b, and c display the phase-averaged tomograms for the mechanical oscillator when prepared in the ground state ($\ket{0}_\text{m}$) and the single-phonon Fock state ($\ket{1}_\text{m}$). We calculate the state preparation fidelity as $0.964 \pm 0.007$ and $0.711 \pm 0.023$ for these states, respectively. Notably, the Wigner distribution for the $\ket{1}_\text{m}$ Fock state exhibits negativity, confirming the non-classical nature of this motional state. From master equation simulations, we identify qubit decoherence ($T_\text{1,q} = 1.55 \:\mu\text{s}$ and $T_\text{2,q}^* = 1.05 \:\mu\text{s}$) as the primary source of infidelity (see \cref{si_Wigner}).

\section*{Mechanical Decoherence}

\begin{figure*}[t]
    \centering
    \includegraphics[width=\textwidth]{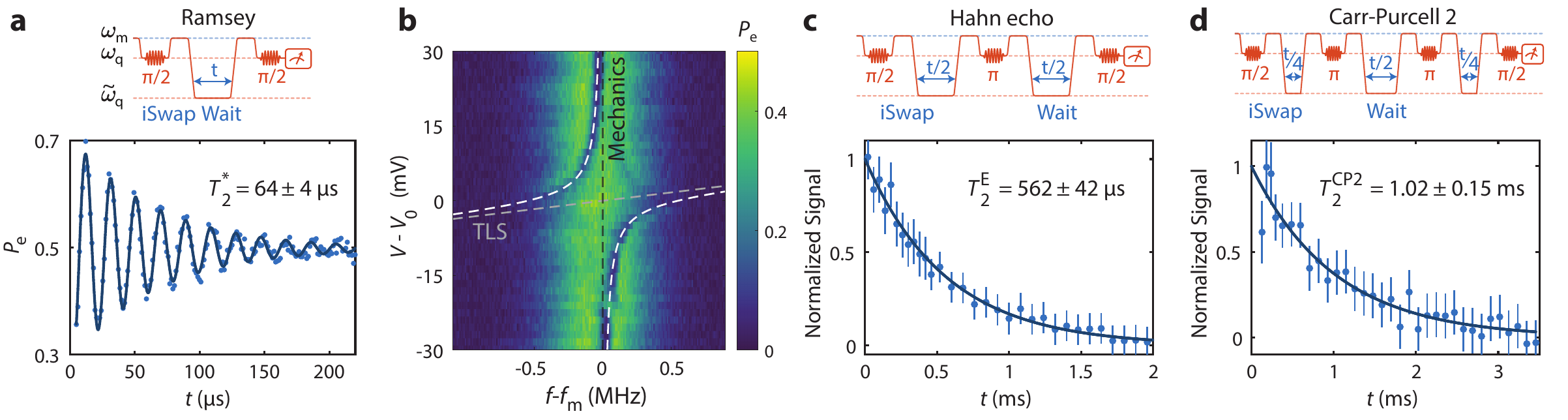}
    \caption{\textbf{Mitigating mechanical dephasing.} \textbf{a,} Ramsey free induction measurement of mechanical coherence. \textbf{b,}  Spectroscopy measurements of the mechanics-qubit system as the voltage is varied, demonstrating an avoided crossing between the mechanical oscillator and a defect whose frequency is shifted with voltage. The white dashed lines are theory fits to the two avoided crossing branches with a coupling rate of $g_\text{TLS}/2\pi \approx 0.5 \text{ MHz}$. The grey dashed line is the bare TLS frequency with a frequency shift rate of $ \approx 0.3 \text{ GHz/V}$. $V_0$ is the voltage at which the TLS and the mechanics are in resonance. Dynamical decoupling procedures in the form of \textbf{c,} Hahn echo measurement with a single refocusing pulse and \textbf{d,} Carr-Purcell (CP) measurement with two refocusing pulses. The error bars represent $1\sigma$ confidence interval.  Dark lines for the coherence measurements (\textbf{a,c} and \textbf{d}) represent theory fits used to extract the coherence times and the top section demonstrates pulse sequences. The measurements are conducted on mechanics A at an applied DC voltage of 50 V.}
    \label{fig:coherence} 
\end{figure*}

Next, we utilize the qubit as a probe for measuring mechanical lifetime. We begin by preparing the oscillator in a single-phonon Fock state ($\ket{1}_\text{m}$, as described previously), allow it to evolve freely for a variable time delay, and measure the remaining population through a swap operation with the qubit. For the oscillator A in our system, we find an intrinsic lifetime of $T_1 = 25 \pm 2 \text{ ms}$ (see \cref{fig:lifetime}). This lifetime is remarkably long, more than two orders of magnitude beyond state-of-the-art circuit quantum acoustodynamics (cQAD) systems \cite{wollack2022a,yang2024,bienfait2019}. We further investigate the mechanical lifetime in our platform by conducting experiments on different devices and across a wide range of phonon numbers, finding consistent results in the same vicinity (see \cref{fig:lifetime}). Remarkably, the energy relaxation dynamics can be well explained by a single exponential decay, with no evidence of commonly observed degradation in the single-phonon regime \cite{wollack2021b,clelandStudyingPhononCoherence2024,emser2024}.
We attribute the ultralong lifetimes and the absence of saturable loss to a low density of defects in crystalline silicon \cite{kleiman1987} and the suppression of phononic density of states in our device geometry \cite{behunin2016}  (see the analysis in \cref{si_acoustic_loss}).

Preserving phase coherence, in addition to minimizing energy decay, is a key requirement for storing quantum information. We characterize mechanical coherence within the Hilbert space spanned by the states $\ket{0}_\text{m}$ and $\ket{1}_\text{m}$ through a Ramsey sequence. In this experiment, the mechanical oscillator is initially prepared in a superposition state $(\ket{0}_\text{m}+\ket{1}_\text{m})/\sqrt{2}$ and allowed to evolve freely for a variable delay, before applying a $\pi/2$ basis rotation and population readout (preparation, manipulation, and readout performed via the qubit; see \cref{fig:coherence}a). We observe that the Ramsey interferogram from this experiment decays exponentially with a time constant of $T_2^* = 64\pm 4 \:\mu\text{s}$. Although this coherence time is comparable to results in other mechanical platforms \cite{yang2024,vonlupkeParityMeasurementStrong2022}, it is considerably shorter than the oscillator's measured lifetime in our system, pointing to significant dephasing.

Similar disparities between energy decay and coherence times have previously been observed in cavity optomechanics \cite{maccabe2020,wallucks2020}, with interactions with TLS defects hypothesized as the underlying cause. We search for signatures of TLS by monitoring the mechanical resonance (via qubit spectroscopy) as we change the voltage bias. Defects within the mechanical oscillators are expected to undergo Stark shift through interactions with the electrostatic field, enabling in-situ modification of the TLS spectral landscape \cite{lisenfeld2019, sarabi2016}. Remarkably, these measurements yield a discrete set of avoided mode crossings that collectively paint the picture of a sparse TLS bath (see \cref{fig:coherence}b). From these measurements, we extract the spectral density and the statistical distributions of the deformation potential and dipole orientations, finding them to be in general agreement with the standard tunneling model \cite{phillips1987} (see \cref{si_tls_spec}). Notably, we observe pronounced modifications of the mechanical frequency noise in the spectral vicinity of the identified defects, displaying the `telegrapher' noise from individual defects (see \cref{{si_freq_noise}}). Collectively, these observations provide experimental evidence for the key components of the TLS-induced decoherence model.

Unlike energy loss, dephasing is, in principle, reversible by reversing a system's time evolution. Such dynamical decoupling operations \cite{carr1954} often require a nonlinear response, which in our setup is provided by coupling to the qubit. \Cref{fig:coherence}c illustrates the control pulse sequence and the measured decay profile from such an experiment, where a refocusing pulse sequence (Hahn echo) is used to mitigate the frequency noise. As evident, the echo coherence time $T_2^\text{E} = 562\pm 42 \:\mu\text{s}$ is substantially longer than $T_2^*$ indicating a ``slow" dominant noise source \cite{ithier2005}. Interestingly, we observe quantitative agreement between the measured improvement from the echo sequence and the predictions from the TLS noise model, suggesting that TLS defects are the dominant source of dephasing (see \cref{si_dephasing}). Building on this approach, we implement a longer echo sequence (Carr-Purcell-2), further enhancing the coherence time to the millisecond level, $T_2^\text{CP2} = 1.02\pm 0.15 \:\text{ms}$ (see \cref{fig:coherence}d). With improvements in the fidelity of qubit-swap operations, even longer echo sequences can potentially extend the coherence time to the limits set by energy decay \cite{bylander2011}.

\section*{Discussion and outlook}
Looking ahead, a natural next step would be attaining the strong dispersive regime, where universal quantum control can be extended to the large bosonic Hilbert space of the mechanical oscillator \cite{10.1038/s41567-022-01630-y}. Achieving this goal requires significant improvements in the electromechanical interaction, which we anticipate to be within reach by miniaturizing the capacitor's vacuum gap \cite{lewis2019} and using superconducting qubits with large flux fluctuations \cite{najera-santos2024,lee2023a}. Alternatively, increased qubit coherence—achievable through flip-chip architectures \cite{rosenberg2017a,conner2021}—can relax the need for stronger interactions. Achieving operation in the strong dispersive regime would enable the encoding of bosonic qubits with biased noise on GHz-frequency mechanical oscillators, potentially leading to scalable architectures for fault-tolerant quantum computation \cite{chamberland2022,guillaud2019,tuckett2018}. Finally, we envision extending the concepts presented here to a wider range of material platforms to probe the ultimate limits of mechanical coherence and to realize superconducting qubit-spin interfaces \cite{neuman2021}.\\

\textbf{Data availability} 
The datasets generated during and/or analyzed during the current study are available from the corresponding author (M.M.) upon reasonable request.
 \begin{acknowledgments}
We acknowledge O. Painter, M. Kalaee, H. Zhao, C. Joshi, F. Yang, and P. Shah for the helpful discussions. This work was supported by the AFOSR (award no. FA9550-23-1-0062) and the NSF (award no. 2137776). A.B. gratefully acknowledges support from the Eddleman Graduate Fellowship. H.T. gratefully acknowledges support from an IQIM Postdoctoral Fellowship. 
\\
\\
\textbf{Author contributions} 
 M.M., A.B., O.G. conceived and designed the experiment. Y.Y. performed numerical optimization of the devices. A.B., O.G., and H.T. worked on the fabrication of the devices. A.B., O.G., and M.M. conducted the measurements and analyzed the data. A.B., O.G., and M.M. wrote the manuscript with input from all authors. M.M. supervised the project.
\\
\\

\end{acknowledgments}
\clearpage
\appendix
\onecolumngrid
\renewcommand{\thefigure}{S\arabic{figure}}
\setcounter{figure}{0}
\renewcommand{\thetable}{S\Roman{table}}
\setcounter{table}{0}

\section{Methods}
\subsection{Summary of device parameters}
The experimentally characterized parameters of our device composed of the qubit, mechanical oscillators, and the readout resonator are listed in \cref{table:param}.
\begin{table}[h]

\begin{center}
\begin{tabular}{c c c c}

Parameter & Value & Description & Component \\
\hline
$\omega_\text{q}^\text{max}/2\pi$ & 5.1071 GHz & maximum frequency & transmon\\
$\alpha_\text{q}/2\pi$ & -226 MHz & anharmonicity & transmon\\
$\omega_\text{r}/2\pi$ & 7.283 GHz & resonance frequency & readout resonator\\
$\chi_\text{qr}/2\pi$ & 450 kHz &  dispersive shift & transmon-readout resonator\\
$\kappa_\text{e,q}/2\pi$ & 7 kHz & external decay rate & transmon\\
$\kappa_\text{e,r}/2\pi$ & 1.9 MHz & external decay rate & readout resonator\\
$\omega_\text{m}/2\pi$ &  4.9176/4.7667 GHz & resonance frequency & mechanics A/B  \\
$g_\text{em}/2\pi$ &  200/230 kHz &  electromechanical interaction rate & mechanics A/B \\
$T_\text{1,q}$ &  1.70/1.55 $\mu$s &  lifetime  & transmon \\
$T_\text{2,q}^*$ &  1.30/1.05 $\mu$s & Ramsey coherence time & transmon \\
$T_\text{1,m}$ &  $19.3\pm1.2/21.1\pm1.1$ ms &  lifetime & mechanics A/B \\
$T_\text{2,m}^*$ &  $64\pm4/67\pm5$ $\mu$s &  Ramsey coherence time & mechanics A/B \\
$T_\text{2,m}^\text{E}$ &  $562\pm42/268\pm23$ $\mu$s & Echo coherence time & mechanics A/B \\
$T_\text{2,m}^\text{CP2}$ &  $1.02\pm0.15/0.54\pm0.11$ ms & CP2 coherence time & mechanics A/B \\
$C_{T_1}$ &  $1.3\times10^5/1.7\times10^5$ & damping cooperativity &  mechanics A/B \\
$C_{T_2}$ &  $132/147$ & dephasing cooperativity &  mechanics A/B \\

\hline\hline
\end{tabular}
\end{center}

\caption{\textbf{Summary of device parameters. } Parameters $\kappa_\text{e,q}$ and $\kappa_\text{e,q}$ denote the decay rates of the qubit and the readout resonator into the shared xy and readout line, respectively. The reported electromechanical coupling rate is characterized at a voltage bias of 50 V. The qubit's lifetime and Ramsey coherence time are characterized and reported at the frequencies of oscillator A/B, respectively. The mechanical lifetime (coherence time) is characterized at a bias voltage of $\approx 40 \text{ V}$ ($\approx 50 \text{ V}$), where the uncertainty in characterization is indicated with $2\sigma$ confidence intervals. The damping (dephasing) cooperativity is calculated as $C_{T_1} = 4g_\text{em}^2T_\text{1,m}T_\text{1,q}$ ($C_{T_2} = g_\text{em}^2T_\text{2,m}^*T_\text{2,q}^*$).}
\label{table:param}

\end{table}

\subsection{Measurement setup}
\begin{figure}[ht]
    \centering
    \includegraphics[width=0.9\textwidth]{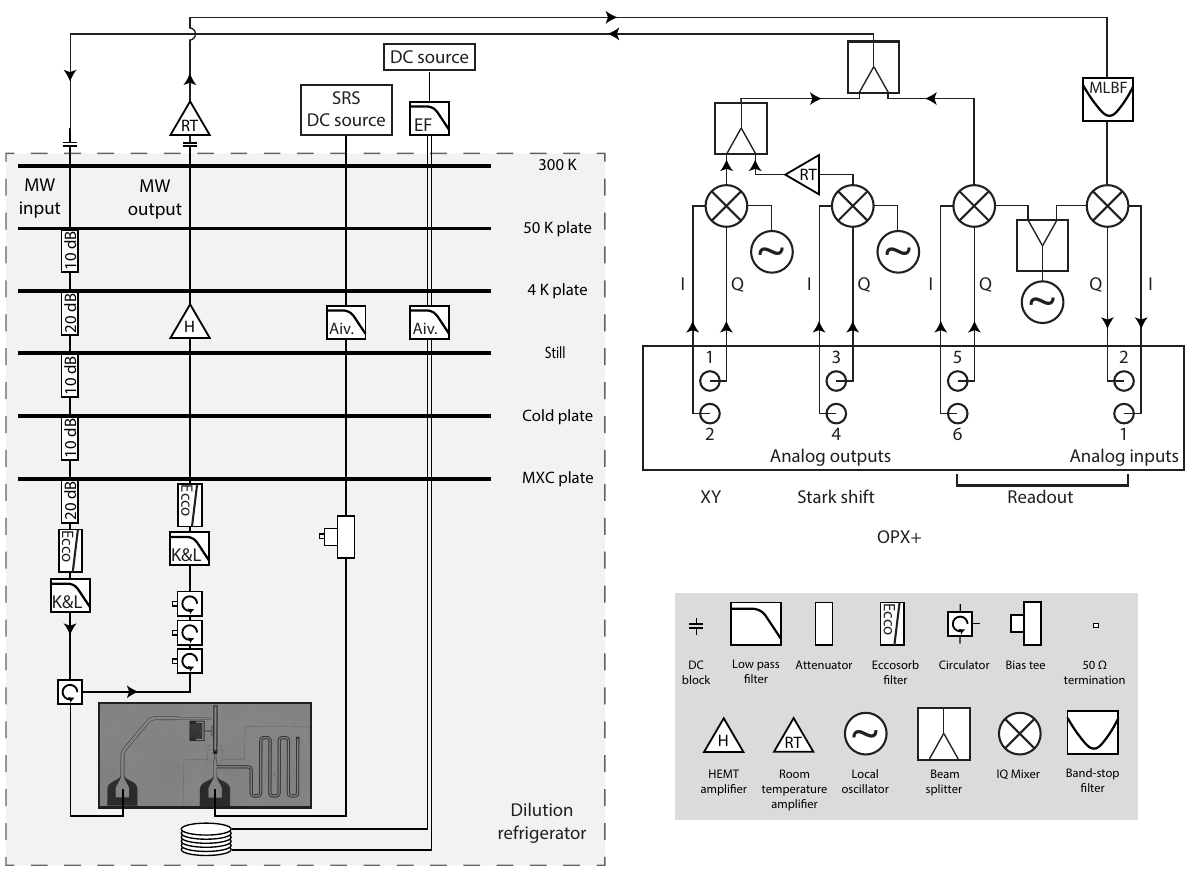}
    \caption{\textbf{The measurement setup.} The schematic of the dilution fridge cryogenic wiring and room temperature electronics.}
    \label{fig:Supp_meas_setup}
\end{figure}

The fabricated chip hosting the device is wire-bonded to a printed circuit board (PCB) and placed in a copper box with a superconducting coil attached to the lid of the box. This coil is used for flux tuning the qubit, where a low-noise DC source (QDevil QDAC) provides the bias current. To further reduce the current noise on the coil, a low-pass electrical filter with a 1 kHz cutoff frequency is employed along with an Aivon Therma-uD25-GL cryogenic RC lowpass filter that has a 30 kHz cutoff frequency. The box, including the chip, is mounted to the mixing stage of a He-3/He-4 dilution fridge and is covered by magnetic shields. The mixing stage of the dilution fridge is cooled down to a base temperature of around 10 mK. The wire delivering the input RF signal to the device is attenuated at each stage of the fridge, as shown in \cref{fig:Supp_meas_setup}, to reduce the thermal noise. Additionally, Eccosorb filters are used to suppress IR radiation, and K$\&$L microwave filters (Tubular .250 Inch Lowpass Filters) with a cutoff frequency of 12 GHz are employed to suppress millimeter-wave radiation. The reflected signal from the device is sent to an output line, where it is amplified using a High-Electron Mobility Transistor amplifier (LNF - LNC0.3-14B HEMT at the 4K stage) and a pair of room-temperature amplifiers. The chip is isolated from the amplifier noise in this path by using an array of three 
 cascaded isolators.

Signal generation and measurements are performed with a multichannel arbitrary waveform generator (AWG) and digitizer (Quantum Machines OPX+). The up- and down-conversion of the signals between the radio frequency and the intermediate frequency band are performed with local oscillators and IQ mixers. For tuning the qubit via the ac Stark shift, we use the up-converted signal from an AWG channel amplified by a room-temperature amplifier. When applying this strong tone, we use a tunable notch filter (Micro Lambda Wireless, MLBFR-0212) in the detection path to avoid saturation at the digitizer.

\subsection{Fabrication}
\label{app:sup_fab}
The devices are fabricated on a 10 mm $\times$ 10 mm, high-resistivity ($> 3 \text{ k}\Omega.\text{cm}$) silicon-on-insulator (SOI) chip with a device layer thickness of 220 nm \cite{kellerTransmonQubitsSiliconinsulator2017a}. The fabrication process consists of the following steps:
(i) Etching of the silicon device layer using Inductively Coupled Plasma Reactive Ion Etching (ICP-RIE) with the $\text{C}_4\text{F}_8/\text{S}\text{F}_6$ pseudo-Bosch process, to define mechanical oscillators, phononic shields, and release holes.
(ii) Electron-beam evaporation of aluminum for the ground plane, DC line, electrodes of moving capacitors, and filter.
(iii) Aluminum evaporation to form the readout line, readout resonator, and transmon capacitor, following a brief vapor hydrofluoric acid (VHF) etch.
(iv) Doubled-angle aluminum evaporation and oxidation to form the Josephson junctions, preceded by a short VHF etch.
(v) Ion milling to remove the oxide on aluminum, followed by aluminum bandage evaporation.
(vi) A longer VHF etching to release the buried oxide layer which partially suspends the device. All the patterning of steps (i)–(v) is carried out using electron-beam lithography.

\section{Electromechanical interaction}
\subsection{System Hamiltonian}
\label{app:sup_hamiltonian}
The Hamiltonian of our system can be written as $\hat{H} = \hat{H}_\text{m} + \hat{H}_\text{q}+ \hat{H}_\text{int}$. Here
\begin{equation}
    \hat{H}_\text{m}/\hbar = \omega_\text{m}\hat{b}^{\dagger}\hat{b},
\end{equation}
is the Hamiltonian of the mechanical oscillator with frequency of $\omega_\text{m}$ and annihilation operator $\hat{b}$. The Hamiltonian of the qubit is 
\begin{equation}
    \hat{H}_\text{q} = 4E_\text{C}\hat{n}^2 - E_\text{J}\cos{\hat{\varphi}},
\end{equation}
where $\hat{n}$ is the charge number operator, $E_\text{C} = e^2/2C_\Sigma$ is the charging energy, $\hat{\varphi}$ is the flux operator, $E_\text{J}$ is the Josephson energy and we have ignored the offset charge due to being in the transmon regime. The interaction Hamiltonian between the mechanical oscillator associated with a moving capacitor $C_\text{m} (x)$ (where $x$ is displacement) and the transmon qubit with total capacitance $C_\Sigma = C_\text{m} + C_\text{q}$ (where $C_\text{q}$ is the qubit capacitance) can be obtained as \cite{lehnert2014}  
\begin{equation}
    \hat{H}_\text{int} = \frac{\hat{q}^2}{2C_\Sigma} \left({\frac{1}{C_\Sigma}\frac{\partial C_{\text{m}}}{\partial x}\hat{x}}\right),
\end{equation}
where $\hat{q} =  2e\hat{n}+Q_\text{dc}$ is the combination of the oscillating RF charge associated with the qubit and the electrostatic charge placed by the external voltage source on the capacitor. The displacement operator can be expressed $\hat{x} = x_{\text{zpf}}(\hat{b} + \hat{b}^\dagger)$, where $x_{\text{zpf}}$ is the zero-point fluctuations of displacement. Substituting these terms into the Hamiltonian, we obtain 
\begin{equation}
    \hat{H}_\text{int} = \frac{1}{2} \frac{\partial C_{\text{m}}}{\partial x} x_{\text{zpf}}  \left(\frac{2e\hat{n}}{C_\Sigma} + V_\text{dc}\right)^2(\hat{b} + \hat{b}^\dagger), 
\end{equation}
where we have used $Q_\text{dc}/C_\Sigma = V_\text{dc}$, with $V_\text{dc}$ being the bias voltage. Expanding the Hamiltonian and keeping only the terms related to the interaction between the qubit and the mechanical oscillator, 
\begin{equation}
    \hat{H}_\text{int} =  \frac{\partial C_{\text{m}}}{\partial x}  x_{\text{zpf}} V_\text{dc} \frac{2e\hat{n}}{C_\Sigma}(\hat{b} + \hat{b}^\dagger). 
\end{equation}
Introducing the resolution of identity for the qubit ($I = \sum_j \ket{j}\bra{j}$) into the equation, the Hamiltonian can be expressed as
\begin{equation}
    \hat{H}_\text{int} =  \frac{\partial C_{\text{m}}}{\partial x}  x_{\text{zpf}} V_\text{dc} \frac{2e}{C_\Sigma} \sum_j\sum_k \ket{j}\bra{j}\hat{n}\ket{k}(\hat{b} + \hat{b}^\dagger)\bra{k}. 
\end{equation}
We can truncate the qubit Hilbert space to the first two levels (ground state $\ket{g}$ and first excited state $\ket{e}$), where the qubit Hamiltonian can be expressed as \begin{equation}
    \hat{H}_\text{q}/\hbar = \frac{\omega_\text{q}}{2}\hat{\sigma}_\text{z},
\end{equation} with $\omega_\text{q}$ corresponding to the frequency difference between the first two energy levels.  After the truncation, the interaction Hamiltonian can be simplified to its Jaynes-Cummings form following a rotating wave approximation, 
\begin{equation}
    \hat{H}_\text{int} =  \frac{\partial C_{\text{m}}}{\partial x}  x_{\text{zpf}} V_\text{dc} \frac{2en_\text{eg}}{C_\Sigma} \left(\sigma_+\hat{b} + \sigma_-\hat{b}^\dagger\right) = \hbar g_\text{em} \left(\sigma_+\hat{b} + \sigma_-\hat{b}^\dagger\right), 
\end{equation}
where $n_\text{eg} = \bra{e}\hat{n}\ket{g}$ is the charge matrix element for the g-e transition, $\sigma_+ = \ket{e}\bra{g}$ ($\sigma_- =\ket{g}\bra{e} $) is the raising (lowering) operator for the qubit. The electromechanical coupling rate is therefore given as 
\begin{equation}    
    \hbar g_\text{em} = \frac{\partial C_{\text{m}}}{\partial x}  x_{\text{zpf}} V_\text{dc} \frac{2en_\text{eg}}{C_\Sigma} = \hbar g_0 V_\text{dc},
    \label{eq:hamiltonian_g}
\end{equation}
where $g_0$ is the coupling rate per 1 V of voltage bias.
For a transmon, the linearized model provides an accurate asymptotic description of the nearest neighbor transition matrix elements \cite{koch2007}, where 
\begin{equation}
    \frac{2en_\text{eg}}{C_\Sigma} \approx \sqrt{\frac{\hbar\omega_\text{q}}{2C_\Sigma}} =  \omega_\text{q} \sqrt{\frac{\hbar Z}{2}} = V_\text{zpf}, 
\end{equation}
with the transmon impedance given as $Z = \sqrt{L_J/C_\Sigma} = \frac{1}{\pi} \frac{h}{(2e)^2} \sqrt{\frac{2E_\text{C}}{E_\text{J}}}$ and $V_\text{zpf}$ corresponding to the zero-point fluctuations of voltage for a linear resonator with impedance $Z$. 

\subsection{Electrical Equivalent Circuit}
\begin{figure}[th]
    \centering
    \includegraphics[width=0.7\textwidth]{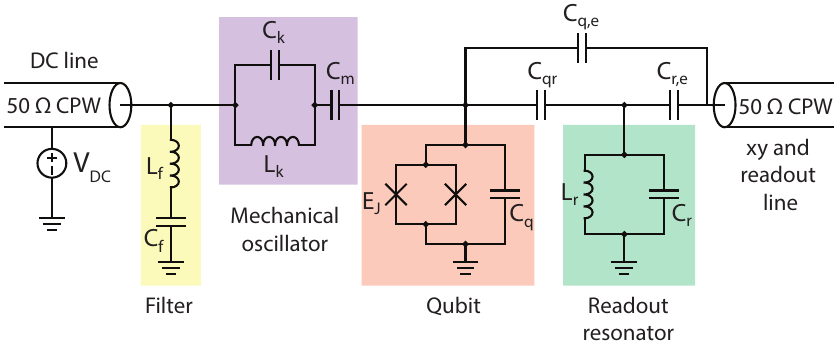}
    \caption{\textbf{Electrical circuit diagram.} The electrical circuit diagram of our quantum acoustics system. The equivalent electrical circuit of the mechanical oscillator is indicated by purple shading.}
    \label{fig:sup_circuit}
\end{figure}
The derived system Hamiltonian which includes the interaction term can be utilized to accurately describe the dynamics of qubit-mechanics interaction. Nonetheless, the analysis of more sophisticated electrical circuits incorporating our mechanical oscillators can benefit significantly from an equivalent electrical circuit for our mechanical circuit, shown in \cref{fig:sup_circuit}. In this equivalent circuit model, $C_\text{m}$ corresponds to the moving capacitor and $C_\text{k}$ is an equivalent capacitance that is given as \cite{bozkurt2023}
\begin{equation}
        C_{\text{k}} = \frac{\hbar C_{\text{m}}^2\omega_m}{2 \left(\partial_x C_\text{m}\, x_\text{zpf}\,V_\text{dc}\right)^2} - C_{\text{m}}.
        \label{eq:equivalent_C}
\end{equation}
The inductance value can be obtained by satisfying the mechanical resonance frequency condition \begin{equation}
    L_\text{k} = \frac{1}{\omega_\text{m}^2(C_\text{k}+C_\text{m})}.
\end{equation}

As can be seen from \cref{fig:sup_circuit}, the interaction with the qubit in this model is due to capacitive coupling, through $C_\text{m}$, which can be expressed as \cite{girvin2014}
\begin{equation}
    \hbar g_\text{em} = \beta \frac{2en_\text{eg}}{C_\text{m}+C_\text{k}} Q_\text{zpf,m},
    \label{eq:capacitive}
\end{equation}
where $\beta = C_\text{m}/C_\Sigma$ and $Q_\text{zpf,m} = \sqrt{\hbar\omega_\text{m} (C_\text{m}+C_\text{k})/2}$ is the zero-point-fluctuations of charge of the electrical equivalent circuit for the mechanical oscillator. If we combine \cref{eq:capacitive,eq:equivalent_C}, we arrive at the expression for the electromechanical coupling rate that was derived based on the interaction Hamiltonian in \cref{eq:hamiltonian_g}. This demonstrates the utility of the electrical equivalent circuit for analysis.

The value of the moving capacitor can be calculated to be $C_\text{m} \approx  0.15 \text{ fF}$, based on finite-element method simulations. At a voltage bias of 50 V, the equivalent electrical circuit parameters are $C_\text{k} \approx 40  \text{ pF}$ (29 pF) and $L_\text{k} \approx 26  \text{ pH}$ (38 pH) for mechanical oscillator A (B) based on the measured electromechanical coupling rate. For different voltages, the equivalent parameters can be calculated by the following scaling relations: $C_\text{k} \propto V_\text{dc}^{-2}$, $L_\text{k} \propto V_\text{dc}^{2}$, which are valid when $C_\text{k}\gg C_\text{m}$.

\section{Mechanical oscillator design}
\label{si_mech_design}
\begin{figure}[t]
    \centering
    \includegraphics[width=0.85\textwidth]{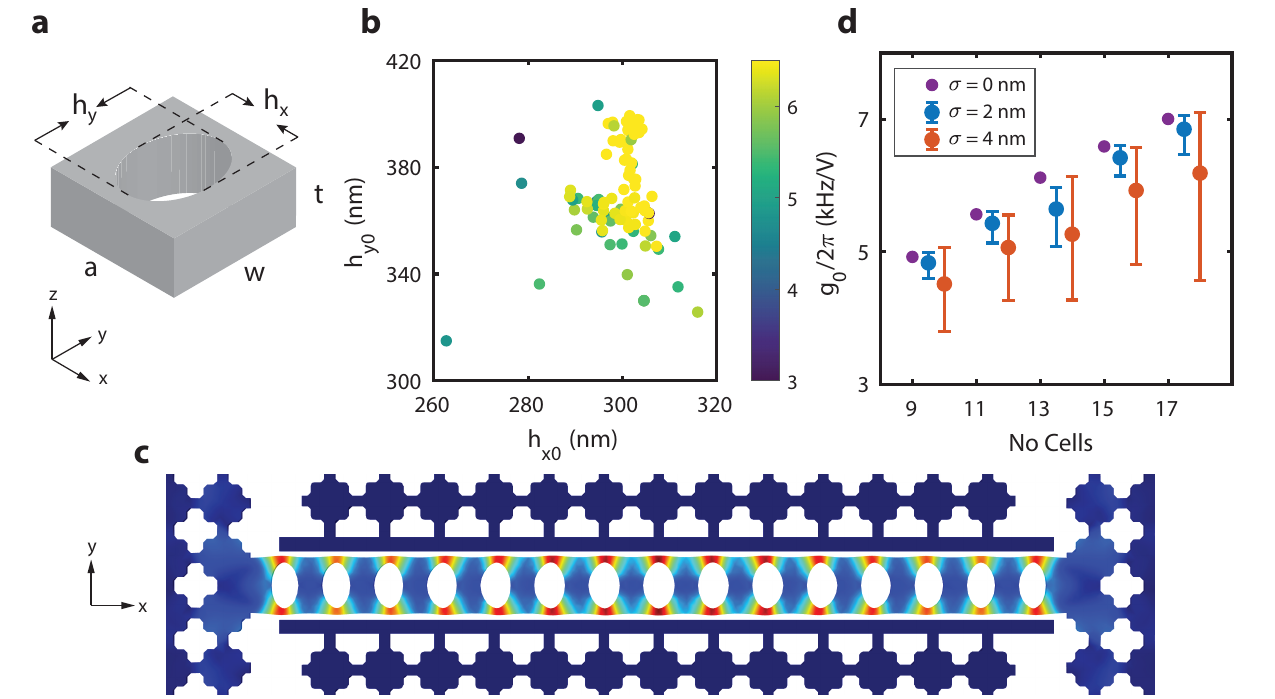}
    \caption{\textbf{Mechanical oscillator design optimization.} \textbf{a,} Phononic crystal unit cell with relevant parameters. \textbf{b,} Optimization results for 15 unit cells, where the electromechanical coupling rate is demonstrated with respect to the elliptical hole dimensions of the central unit cell for each iteration. The convergence takes $\sim 100$ iterations. \textbf{c,} Simulated exaggerated mechanical displacement for the optimized structure. \textbf{d,} Electromechanical coupling rate statistics obtained from Monte-Carlo simulations with varying degrees of disorder for different numbers of unit cells. The legend represents the disorder magnitude $\sigma$. Zero disorder corresponds to ideal disorder-free devices. Each point corresponds to 50 different disorder realizations and the error bars indicate the two standard deviation confidence interval. The data with disorder is offset horizontally from the disorder-free points for visualization.    }
    \label{fig:sup_mech_des}
\end{figure}

The removal of the metal from the central nanobeam of our phononic crystal transducer comes with a significant reduction (approximately a factor of 6) in the electromechanical coupling rate. Combined with the utilization of a relatively low impedance ($ Z = 350 \:\Omega$) microwave qubit in the form of a charge-insensitive transmon, potential avenues to boost the electromechanical coupling rate become crucial. To this end, we numerically optimize our phononic crystal resonator, whose unit cell is depicted in \cref{fig:sup_mech_des}a, following previous methods to optimize optomechanical crystal cavities \cite{chan2012,moraes2022a}. For the optimization procedure, the cost function is defined to be $-g_\text{em}$, which is minimized by the Nelder-Mead algorithm \cite{olsson1975}. The result of an example optimization run is visualized in \cref{fig:sup_mech_des}b, demonstrating the improvement optimization provides. The optimized structure possesses a uniform displacement profile throughout the entire beam, as depicted in \cref{fig:sup_mech_des}c. This is in line with the analytical expression for the electromechanical coupling rate \cite{bozkurt2023}, which suggests that the coupling rate is maximized for a uniform displacement profile.

Though the optimization procedure leads to an improvement in the electromechanical coupling rate for an ideal structure, the physically attainable coupling rates and cavity sizes are set by fabrication disorder. The impact of fabrication disorder on devices can be numerically modeled through Monte Carlo simulations \cite{minkov2013}.  Via this method, we investigate the impact of varying degrees of disorder on the electromechanical interaction rate, as shown in \cref{fig:sup_mech_des}d. We observe that for lower levels of disorder, there is a steady improvement in the coupling rate with an increased number of cells, whereas the improvement starts to reduce for a higher degree of disorder. Based on the characterization of previously fabricated devices, we expect our fabrication disorder to be at the level of $\sigma = 2-4\text{ nm}$ \cite{bozkurt2023}. Relying on these simulations, we have fabricated cavities with 15 cells. We have measured $g_0 = 4.2 \text{ kHz/V}$, which is in reasonable agreement with the 4 nm disorder result of $5.9^{+0.6}_{-1.2} \text{ kHz/V}$. This is potentially indicative of a relatively high level of disorder in our structures. Methods to mitigate disorder in order to further boost interaction rates will be the subject of future studies.
\begin{figure}[hbt]
    \centering
    \includegraphics[width=0.5\textwidth]{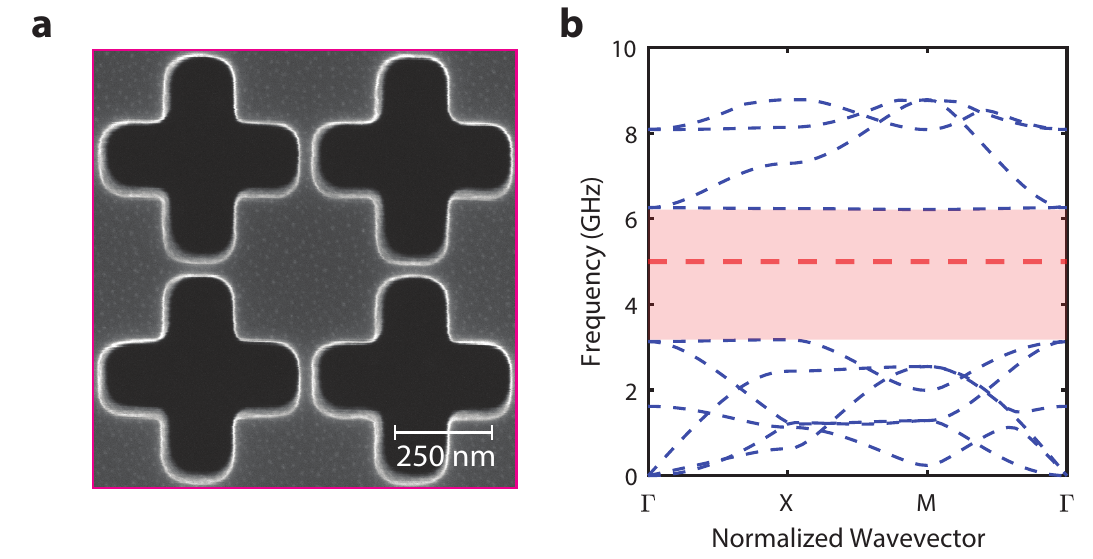}
    \caption{\textbf{Phononic shields.} \textbf{a,} Scanning electrode microscope image of cross-type phononic shields. \textbf{b,} Acoustic band diagram of the phononic shield unit cell. The shaded region represents the bandgap and the red dashed horizontal line indicates a frequency of 5 GHz.}
    \label{fig:sup_mech_shield}
\end{figure}

All of the mechanical resonators are clamped by identical phononic shields \cite{chan2012,maccabe2020}, which are displayed in \cref{fig:sup_mech_shield}a. This is possible due to the wide ($>2 \text{ GHz}$) full bandgap of the phononic shields (see \cref{fig:sup_mech_shield}b). The effectiveness of phononic shields in suppressing acoustic radiation has been shown to depend on fabrication disorder \cite{maccabe2020}. The impact of the fabrication disorder on acoustic radiation can be estimated through Monte Carlo simulations, similar to our examination of the effect of disorder on interaction rate. Our shielding structure that has 8 unit cells is expected to limit radiation to a small rate of $\approx 2\pi\times 1 \text{ mHz}$ ($Q > 10^{12}$), based on our numerical simulations with 4 nm of disorder. 

\section{Microwave design}
\label{si_mw_design}
Our microwave qubit is a transmon with $E_\text{J,max}/h = 15.9 \text{ GHz}$ and $E_\text{C}/h = 226 \text{ MHz}$, having a maximum frequency of 5.107 GHz. The transmon is designed to be insensitive to charge noise from the voltage supply. At the lowest frequency of operation, around mechanics B, the charge dispersion is $\approx 2 \text{ kHz}$ for the g-e transition \cite{koch2007}. The SQUID loop of the transmon is comprised of two asymmetric junctions, with an asymmetry ratio of 6. The asymmetric junctions reduce the sensitivity to flux noise as the qubit frequency is tuned at the cost of a reduced tuning range \cite{hutchings2017}, which is $\approx 800 \text{ MHz}$ in our case. However, even with reduced sensitivity to flux noise, we observe that the coherence time of our qubit away from the sweet spot is limited mainly by flux noise with an estimated amplitude of $\approx 10 \,\mu\Phi_0$.

\begin{figure}[h]
    \centering
    \includegraphics[width=0.8\textwidth]{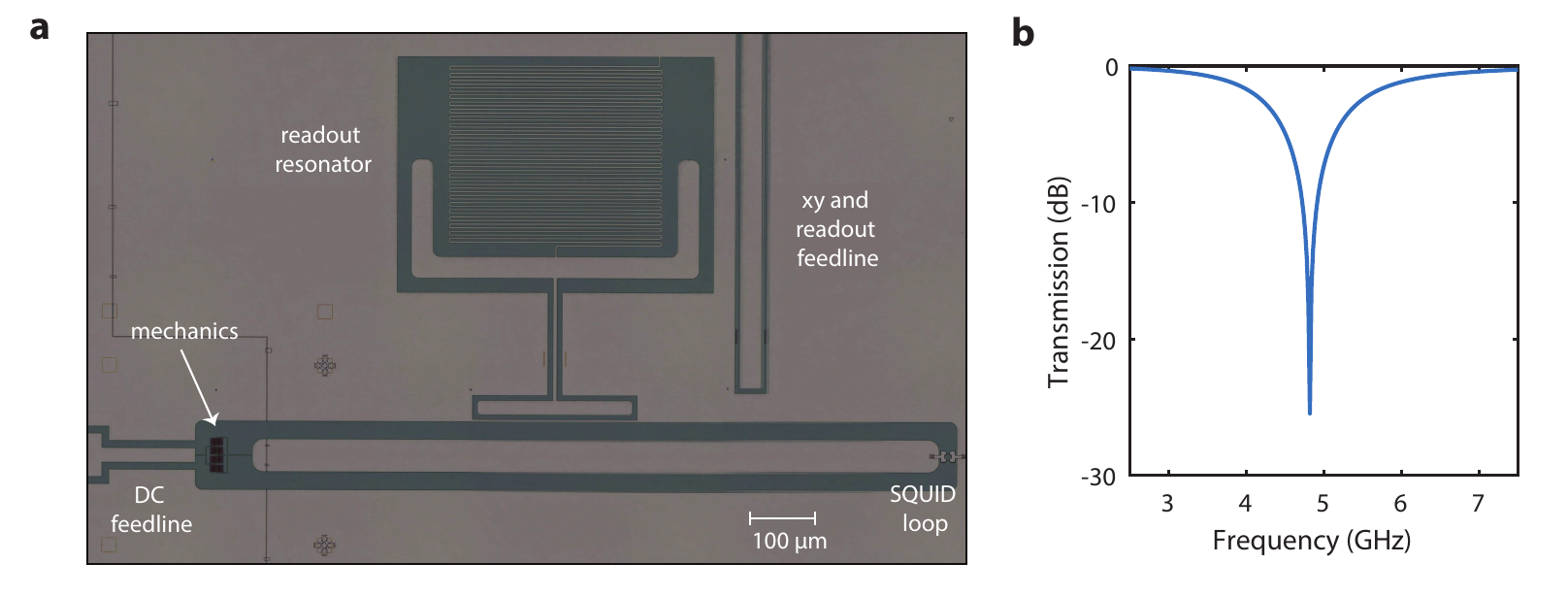}
    \caption{\textbf{Microwave design.} \textbf{a,} Optical microscope image of the device, focusing on the features around the transmon. \textbf{b,} Simulated transmission between the launchpad and the feedline for the DC line, showing a notch type response centered at $4.8 \text{ GHz}$.}
    \label{fig:sup_filter}
\end{figure}

The qubit is capacitively coupled to a lumped element $\lambda/4$ resonator at a frequency of $\omega_\text{r}/2\pi \approx 7.3 \text{ GHz}$ with an  interacton rate $g_\text{qr}/2\pi = 105 \text{ MHz}$ for dispersive readout (see \cref{fig:sup_filter}a). The coplanar waveguide that is capacitively coupled to the readout resonator ($\kappa_\text{e,r}/2\pi = 1.9 \text{ MHz}$) is also coupled to the qubit ($\kappa_\text{e,q}/2\pi = 7 \text{ kHz}$) to function as both a readout and xy line. The xy coupling to the qubit is engineered to be relatively strong in order to reduce power requirements for ac Stark shift tones, which are also applied through the same port.

The mutual capacitance between the DC line and the transmon qubit is $\approx 2 \text{ fF}$ according to electromagnetic simulations. This capacitance consists of the mechanical capacitance, the parasitic capacitance of the wires and the stray capacitance between the feedline and the transmon central electrode. In the absence of any filtering, this undesirable capacitive coupling between the DC line and qubit is expected to cause external decay with a rate of $\approx 2\pi\times 350 \text{ kHz}$ into the waveguide. In order to avoid this, we utilize a notch-type filter. At its resonance frequency, the $\lambda/4$ CPW resonator functions as an impedance transformer, where the $50 \,\Omega$ environment is converted into a short, preventing radiation \cite{reed2010}. The precise design of the filter relies on electromagnetic simulations, where the length of the $\lambda/4$ stub is designed such that the transmission from the launchpad of the waveguide to the feedline that is connected to the mechanical resonator is suppressed at the qubit frequency, as seen in \cref{fig:sup_filter}b. Due to the stub being galvanically attached to the waveguide and being open on the other end, it doesn't interfere with the application of DC voltage. However, the filter also precludes direct electrical addressing of the mechanical resonators via the DC line and instead the driving of mechanical oscillators has to be realized via the xy line, mediated by the qubit. This necessitates the fabrication of other devices on the chip for high phonon number free decay measurements.

\section{Voltage biasing}
\subsection{DC voltage setup}
The DC voltage for electrostatic biasing is applied with a Stanford Research Systems DC205 voltage source, whose power supply is filtered with an Onfilter CleanSweep AC EMI filter. The twisted pair lines that carry the DC voltage to the device are further filtered at the 4 K stage breakout with an Aivon Therma-uD25-GL cryogenic RC lowpass filter that has a 30 kHz cutoff frequency. Finally, at the mixing stage, the Quantum Microwave QMC-CRYOTEE-0.218 bias-T acts as a lowpass filter with a 600 kHz cutoff frequency for the DC bias. A $50 \,\Omega$ termination is attached to the RF port of the bias-T.

\subsection{Impact of voltage bias on qubit}
\label{si_dc_qubit}

\begin{figure}[tbh]
    \centering
    \includegraphics[width=\textwidth]{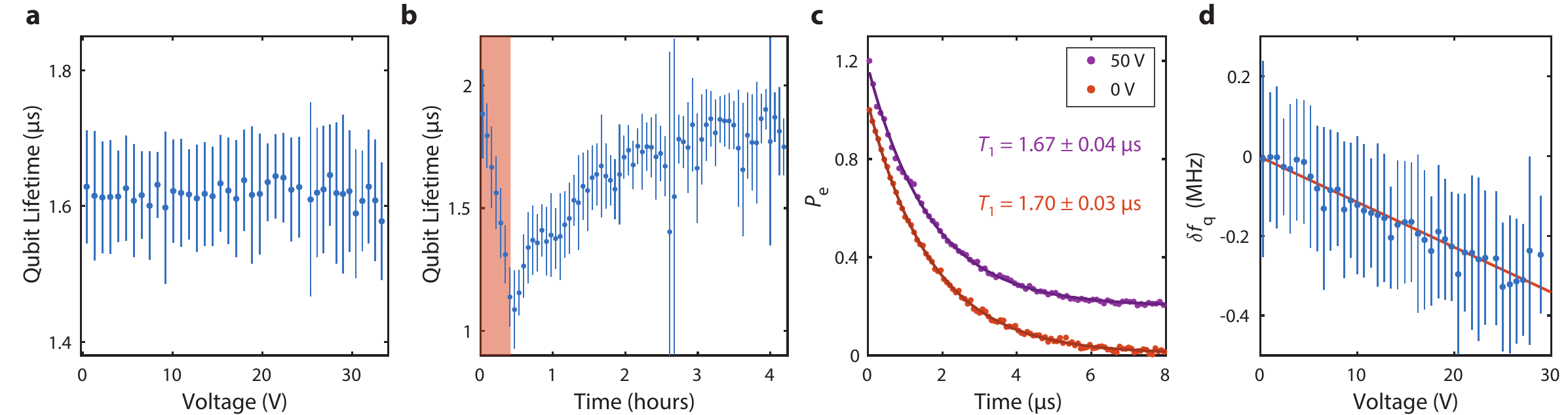}
    \caption{\textbf{DC voltage effects on qubit.} \textbf{a,} Qubit $T_1$ lifetime as the voltage is ramped continuously from 0 V to 25 V with a rate of $2 \text{ mV/s}$ and from 25 V to 35 V with a rate of $1 \text{ mV/s}$. \textbf{b,} Qubit $T_1$ lifetime degrading due to ramping of the voltage and subsequently recovering with time. The red shaded region indicates the duration where the voltage was increased from 37 V to 40 V with a rate of $2 \text{ mV/s}$. \textbf{c,} Qubit decay measurements without any voltage applied and at 50 V following recovery of the qubit. The solid lines are exponential fits. The data is offset vertically for clarity. \textbf{d,} Changes in qubit frequency ($\delta \text{f}_\text{q}$) vs voltage obtained by Ramsey measurements. The red solid line is a linear fit with slope $-11 \text{ kHz/V}$. The lifetime measurements presented in \textbf{a} is interleaved with Ramsey measurements utilized to extract qubit frequency in \textbf{d}. Each data point in \textbf{a,b,d} represents the average of 30 consecutive measurements with error bars indicating $2\sigma$ confidence interval, which is mainly set by time fluctuations of lifetimes and frequencies.}
    \label{fig:sup_volt_qubit}
\end{figure}

Since the DC voltage is applied to the mechanical resonator, which is in turn attached to the qubit, it is crucial to investigate potential parasitic effects on the qubit that may degrade its performance. To this end, we monitor our qubit by conducting decay and Ramsey measurements as the voltage is varied. 
We observe that the qubit lifetime does not exhibit any systematic degradation as the voltage is slowly ramped to around 35 V (see \cref{fig:sup_volt_qubit}a), with a rate of $\sim 1 \text{ mV/s}$, similar to \cite{lahaye2015a}. However, at higher voltages we start to see a transient degradation of the qubit relaxation lifetime as the voltage is increased, followed by a slow recovery on the timescale of hours as seen in \cref{fig:sup_volt_qubit}b. The excess transient decay could be due to a currently unknown mechanism that generates quasiparticles during the charging of our capacitor, however, the long recovery timescale is not consistent with the quasiparticle recombination timescale on aluminum ($\sim 10 \text{ ms}$) \cite{wang2014}. We qualitatively observe that reducing the ramp rate helps reduce the transient degradation of the qubit and the degradation strongly depends on the voltage at which the ramp happens, becoming worse at higher voltages where the recovery takes longer as well. Further studies will be needed to better understand the underlying physical mechanisms. Nonetheless, we observe that the qubit completely recovers following a ramp, a behavior repeatedly verified up to our maximum voltage of 50 V (limited by the bias-T) as seen in \cref{fig:sup_volt_qubit}c. The $T_2^*$ lifetime obtained by Ramsey measurements are identical for 0 V and 50 V as well, testifying to the absence of excess dephasing due to the voltage bias.

Apart from enabling us to explore the voltage dependence of decoherence, Ramsey measurements also provide information about the qubit frequency, based on the frequency of the Ramsey fringes. Interestingly, we observe that the qubit frequency reduces with voltage, as shown in  \cref{fig:sup_volt_qubit}d. Since quasiparticles are known to reduce qubit frequencies, whether the observed reduction is consistent with quasiparticle physics need to be explored. Relying on analytical expressions for the quasiparticle induced frequency shift \cite{catelani2011}, we calculate that for $\approx 500 \text{ kHz}$ frequency shift, the required quasiparticle density is $x_\text{qp}\approx 10^{-4}$. This relatively large quasiparticle density would limit the qubit lifetime to $T_\text{1,q}< 300 \text{ ns}$ and increase the qubit excited state population to $P_\text{e} > 0.1$ \cite{jinThermalResidualExcitedState2015}, inconsistent with the absence of noticeable lifetime degradation and heating we have measured. Therefore, the frequency shift is unlikely to be caused by quasiparticles and can instead be attributed to either the vacuum gaps or the membrane separation of the device shrinking due to electrostatic forces, which in turn increase the total capacitance of the transmon. Similar reductions in frequency has been previously observed with electrostatic transducers attached to linear resonators \cite{bozkurt2023}.

In combination with the thermometry results that show identical qubit thermal occupation at 0 V and at 50 V, our measurements indicate the absence of any noticeable steady-state degradation of our qubit due to the applied voltage.

\subsection{Leakage current measurements}

\begin{figure}[tb]
    \centering
    \includegraphics[width=0.3\textwidth]{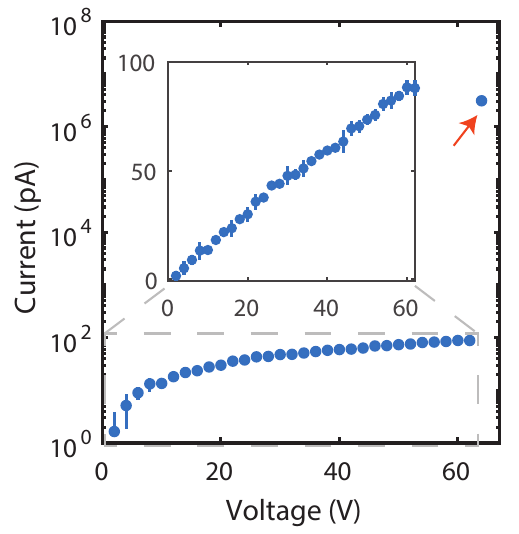}
    \caption{\textbf{Leakage current measurements.}  Leakage current measured by the picoammeter for waveguide device I as the voltage is varied. The inset shows the region from 0 V to 62 V where the current increases linearly. The baseline resistance is set by the cable resistance, which was verified by detaching the device through a switch and observing an unchanged slope for the current. The error bars indicate $2\sigma$ confidence intervals. }
    \label{fig:sup_leak}
\end{figure}
We have discussed in detail the effects of the voltage bias on the qubit, through time domain characterization and thermometry measurements. In our experiments, the maximum voltage we could reach of 50 V was set by the bias-T. However, for future experiments, understanding the maximum voltage that the devices can endure is important, since it provides information about the ultimate limits of coupling enhancement. To explore this question, we have fabricated devices that have a structure comparable to our quantum acoustic system. Importantly, when a voltage is applied to a device, the current supplied by the voltage source (referred to as leakage current) provides information about the onset of breakdown or pull-in events \cite{ritter2021,bozkurt2023}. Since this is an exclusively DC measurement, we can apply a voltage without needing a bias-T and utilize a picoammeter (Keithley 6485) to measure the leakage current. As shown in \cref{fig:sup_leak}, we observe a small current that increases linearly with voltage (attributed to cable resistance) until 64 V, where it abruptly increases, consistent with the onset of pull-in instability \cite{bozkurt2023}. These results indicate that with an improved setup, there may be room to further enhance the electromechanical interaction rate via higher bias voltages.

\section{Thermometry}
\label{si_thermometry}

To do measure the qubit's effective temperature, we use the Rabi Population Measurement (RPM) method, introduced by Geerlings \textit{et al.} \cite{geerlingsDemonstratingDrivenReset2013}. For this purpose, we first conduct two-tone spectroscopy to find the frequency of the qubit's e-f transition. We then measure the baseline or reference population by applying an e-g $\pi$-pulse, followed by an X rotation on the first and second excited states, and do qubit population measurement.
The $\pi$-pulse swaps the ground and first excited state populations, and the X rotation puts the population in the first excited state to a superposition of the first and second excited states. By varying the angle of the X rotation, we get Rabi oscillations where the amplitude of the measured signal is proportional to the initial ground state population $P_g$. To measure the population of the first excited state we repeat the same procedure without the initial $\pi$-pulse, which results in the measured amplitude being proportional to the first excited state population $P_e$. By assuming that the thermal populations in higher excited states are negligible, which is justified based on the resulting low residual thermal population, and normalizing the measured populations to their sum, we get the residual thermal population of the qubit. The amplitude of the oscillations in \cref{fig:Wig} are normalized to this sum, such that the magnitude of the oscillations in the right panel is proportional to the thermal populations.

\begin{figure}[ht]
    \centering
    \includegraphics[width=.3\textwidth]{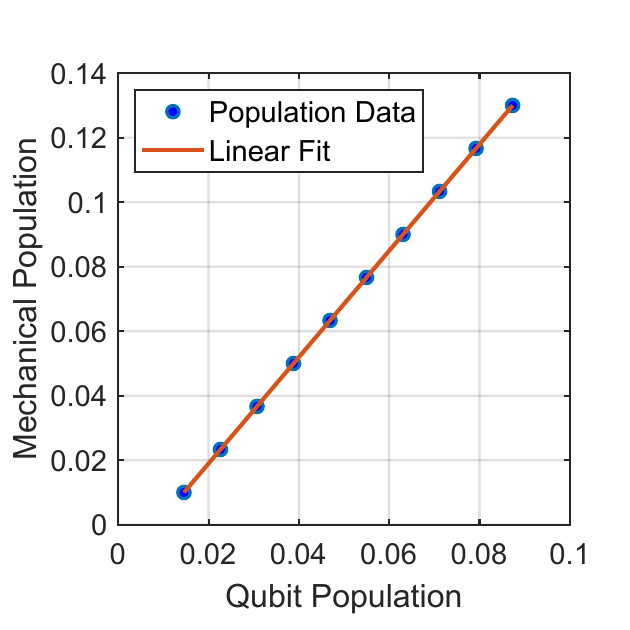}
    \caption{\textbf{Qubit vs mechanics population.} Simulated corresponding mechanics population of mechanics B for a given measured qubit population in the thermometry setup. For the simulation, we start with a qubit thermal population of 4\%, caused by the ac Stark shift, and varying mechanics population. Then we let the qubit and mechanics interact on resonance for half a Rabi cycle and measure the resulting qubit population. Notably, during the qubit-mechanics interaction we use a bath population of 2\% for the qubit dissipator since the Stark drive is turned off for this step.}
    \label{fig:qubit_vs_mech_pop}
\end{figure}

To use the same technique for mechanics thermometry we do a swap operation between the mechanics and qubit before the experiment described above. To ensure the oscillator is in equilibrium with its thermal bath, we wait around 60 ms (which is much longer than the mechanical lifetime) before each measurement.  We use numerical modeling to back out the mechanical temperature from the results of this experiment to properly account for the infidelity of the swap operation, the potentially different temperatures of the qubit's and oscillator's baths, and nonzero population of the qubit caused by the ac Stark shift tone, which is needed to adjust the qubit-mechanics detuning. We measure this ac Stark shift drive-induced population of the qubit to be $0.039 \pm 0.003$ using the qubit thermometry while this drive tone is on and use it in a use Master equation analysis with the QuTiP package \cite{johanssonQuTiPPythonFramework2013, johanssonQuTiPOpensourcePython2012a} to find the corresponding mechanical population (see \cref{fig:qubit_vs_mech_pop}).

\section{Wigner tomography}
\label{si_Wigner}

We calculate the Wigner function of the mechanical oscillator initialized to state $\rho$ by measuring the phonon parity. The Wigner function of the prepared state $\rho$ at displacement $\alpha$ ($W_\rho(\alpha)$) is proportional to the displaced parity of the mechanical oscillator $P_\rho(\alpha)$, through the relation 
\begin{equation}
    W_\rho(\alpha) = \frac{2}{\pi}P_\rho(\alpha).
\end{equation}
To calculate the parity we leverage the information in the Rabi oscillations between the qubit and the mechanical oscillator \cite{hofheinzSynthesizingArbitraryQuantum2009,wollack2022a}. From these Rabi oscillations, we calculate the phonon number distribution, and consequently, phonon parity which are related by
\begin{equation}
    P_\rho(\alpha) = \text{Tr} \left (\hat{D}^\dagger(\alpha)\hat{\rho}\hat{D}(\alpha)\hat{\Pi}\right ) = \sum_n (-1)^n p_{n,\rho}(\alpha),
\end{equation}
where $\hat{\Pi}$ is the parity operator and $p_{n,\rho}(\alpha)$ is the phonon population corresponding to the Fock state n for an initialized state of $\rho$ following a displacement of $\alpha$. 
Importantly, during our measurements, we randomize the phase of the displacement $\alpha = r e^{i\theta}$ (where $\theta$ is sampled uniformly from the interval $[0,2\pi]$), which leads to the measurement of the phase averaged Wigner function $\overline{W}(r)$ \cite{lvovsky2001}. Since the states we are interested in preparing ($\ket{0}_\text{m}$ and $\ket{1}_\text{m}$) are phase-symmetric, the phase averaging provides a significantly reduced measurement overhead, while retaining fundamental features of the Wigner function. 

\begin{figure}[ht]
    \centering
    \includegraphics[width=.8\textwidth]{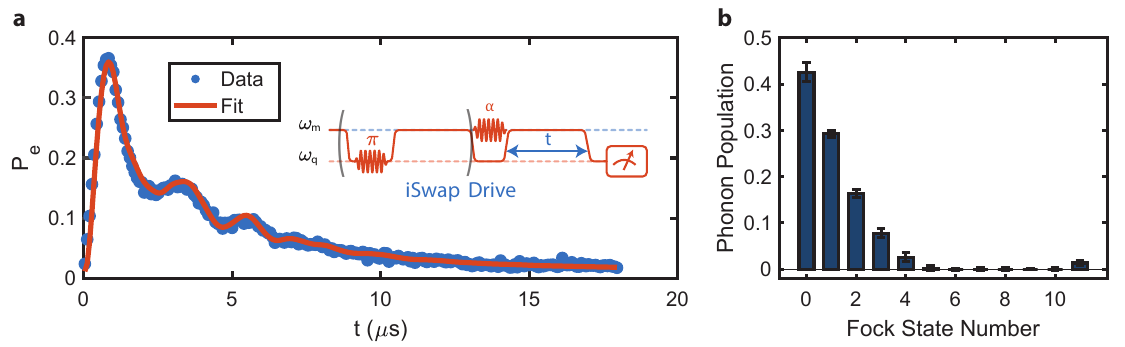}
    \caption{\textbf{Phonon distribution measurement.} \textbf{a,} Measurement of the qubit population as a function of time for the ground state tomography of mechanics B with a displacement pulse of amplitude 0.9. The inset depicts the pulse sequence for the ground (excited) state tomography. \textbf{b,} Calculated phonon distribution from the fit.}
    \label{fig:Supp_wigner}
\end{figure}

To perform Wigner tomography after preparing a mechanical state, we first detune the qubit by approximately 5 MHz and then apply a phase-randomized Gaussian pulse to the qubit at mechanics frequency for $8 \ \mu$s. This drive, mediated via the qubit through the electromechanical interaction, effectively converts into a displacement of the mechanical oscillator with complex displacement amplitude $\alpha$, where we choose the phase of $\alpha$ randomly for each instance of the experiment. Subsequently, we bring the qubit into resonance with mechanics for time t and measure the qubit population as a function of t. This results in Rabi oscillation where an example of such measurement is depicted in \cref{fig:Supp_wigner}a.
We prepare the ground mechanical state by moving the qubit to the mechanics' frequency and waiting approximately $25\ \mu$s to make sure that they are both at their ground states. For the excited state preparation, we repeat the same procedure followed by detuning the qubit by around 5 MHz and exciting it using a $\pi$-pulse. Then, we bring the qubit back into resonance with the mechanics and wait for $T_\text{swap}$ to transfer the excited state to mechanics. The pulse sequences for the ground and excited state preparation and tomographies are shown as the inset of \cref{fig:Supp_wigner}a.

To derive the phonon number distribution, we fit the measured Rabi oscillation using the QuTiP package to solve the master equation. For this purpose, we first find the resulting Rabi oscillations corresponding to mechanics being in the Fock state n and qubit in the ground state ($\ket{g, n}_{q,m}$) \cite{hofheinzSynthesizingArbitraryQuantum2009}. Then, by decomposing the measured Rabi oscillations into Rabi functions from the simulation, the phonon number distribution is calculated. As discussed above the Wigner function is calculated for each $\alpha$, which gives us the phase-averaged measured Wigner function. For the errors, we use bootstrapping done by resampling with replacement from the calculated data. An example of such fitted Rabi oscillation and phonon number distribution is shown in \cref{fig:Supp_wigner}a, and b, respectively. Additionally, to characterize the magnitude of the coherent displacement, we use a Poisson fit to the phonon number distribution for ground state tomography and calibrate the drive amplitude corresponding to a required displacement amplitude.

After measuring the Wigner function for the prepared density matrix $\rho$, we reconstruct this density matrix $\rho'$ of the prepared mechanical states by minimizing the difference between the measured Wigner function and the calculated Winger function given by
\begin{equation}
    W_{\rho'}(\alpha) = \frac{2}{\pi}\sum_n \bra{n}\hat{D}^\dagger(\alpha) \hat{\rho}' \hat{D}(\alpha) \hat{\Pi}\ket{n}.
\end{equation}
For this minimization, we used CVX, a package for solving convex programs \cite{cvx,gb08}, with a Hermitian semi-definite $\rho$ and subjecting it to be diagonal (due to the measurement being phase-randomized and hence the Wigner function being spherically symmetric) and have a trace of one. For the errors, we use the Monte Carlo method where we generate statistics for the elements of the density matrix by choosing a phonon distribution value using Gaussian distribution with the mean and standard deviation of the calculated values and errors in the previous step, respectively.

The expected state preparation fidelities can be estimated through master equation simulations of the qubit-oscillator system. At the measurement voltage of 50 V, the electromechanical interaction rate is given to be $2\pi\times 230 \text{ kHz}$ for oscillator B. The decoherence rates utilized in the simulation is obtained through time domain characterization. Near mechanical oscillator B, the qubit has a lifetime of $T_\text{1,q} = 1.55 \:\mu\text{s}$ and a Ramsey coherence time of $T_\text{2,q}^* = 1.05 \:\mu\text{s}$, constituting the main decoherence channel for the qubit-oscillator system ($T_\text{2,B}^* \approx 50 \:\mu\text{s} \gg T_\text{2,q}$). Further utilizing the thermal populations extracted through thermometry measurements, the state preparation fidelities for $\ket{0}$ and $\ket{1}$ are estimated to be 0.989 and 0.759, comparable to the measured fidelities of $0.964 \pm 0.007$ and $0.711 \pm 0.023$ (where the fidelity is defined as $F = \sqrt{\bra{\psi}\rho\ket{\psi}}$, with $\ket{\psi}$ being the target state). These observations suggest that qubit decoherence is the main contributor to state preparation infidelity, with the remaining infidelity most likely originating from imperfections in experimental procedures. 

\section{Mechanical lifetime measurements}
\label{si_lifetime}

\subsection{The inverse Purcell decay}
\label{si_Purcell}

\begin{figure}[ht]
    \centering
    \includegraphics[width=.3\textwidth]{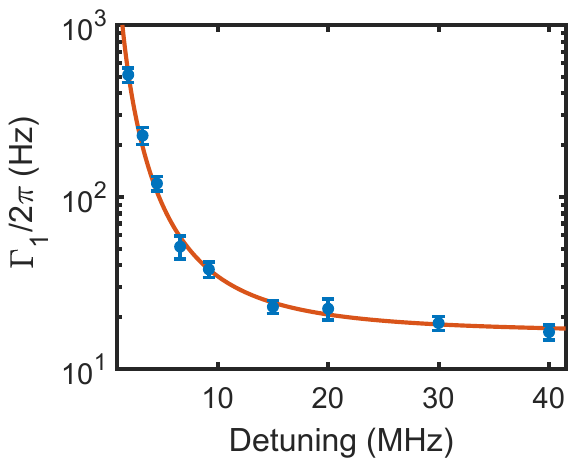}
    \caption{\textbf{Inverse Purcell decay.} Mechanical decay rate as a function of the detuning between the qubit and mechanics, demonstrating the inverse Purcell decay of the mechanics through the qubit, performed on mechanics A. The data points indicate the measured values and the solid line shows the theory fit based on \cref{eq:Supp_Purcell}.
    }
    \label{fig:Supp_Purcell}
\end{figure}

In our qubit-mechanics coupled system, the mechanical oscillators experience inverse Purcell decay due to their interaction with a much shorter-lived qubit, which limits their lifetime, particularly at small detunings. The measured decay rate of the mechanics can therefore be expressed as the sum of intrinsic and inverse Purcell decay rates, $\Gamma = \Gamma_\text{i} + \Gamma_\kappa$. For typical detunings of our experiment, where $ |\Delta| \gg \max(\kappa, g)$, the total decay rate is given by \cite{setePurcellEffectMicrowave2014}
\begin{equation}
    \Gamma = \Gamma_\text{i} + \left(\frac{g_\text{em}}{\Delta}\right)^2\kappa.
    \label{eq:Supp_Purcell}
\end{equation}
Here, $g_\text{em}$ denotes the electromechanical coupling rate, $\Delta = \omega_\text{q} - \omega_\text{m}$ is the detuning between the qubit and the oscillator, and $\kappa$ is the qubit decay rate. To quantify the contribution of the inverse Purcell decay, we measure the mechanical decay rate while varying the detuning $\Delta$. As shown in \cref{fig:Supp_Purcell}, the theoretical model aligns closely with the experimental data. From the fit parameters, we calculate the intrinsic decay rate of the mechanics by subtracting the inverse Purcell decay contribution. These measurements were performed in the low-phonon-number regime, where we subtract a small background offset—accounting for the residual thermal population—from all data points. To minimize the inverse Purcell decay during the free energy decay measurements, we detune the qubit frequency from the mechanical resonance by approximately 30–40 MHz during relatively long waiting times. However, since rapid tuning of the qubit frequency via an ac Stark shift \cite{carrollDynamicsSuperconductingQubit2022} requires a microwave tone, larger frequency shifts require higher drive power, which can induce qubit heating. To address this, we maintain a smaller detuning of around 3–6 MHz during fast operations, such as qubit gate operations and readout, as oscillators are less affected by the inverse Purcell decay.

\subsection{Intermediate-phonon-number measurement}
\label{si_mid_PhN}
\begin{figure}[ht]
    \centering
    \includegraphics[width=.6\textwidth]{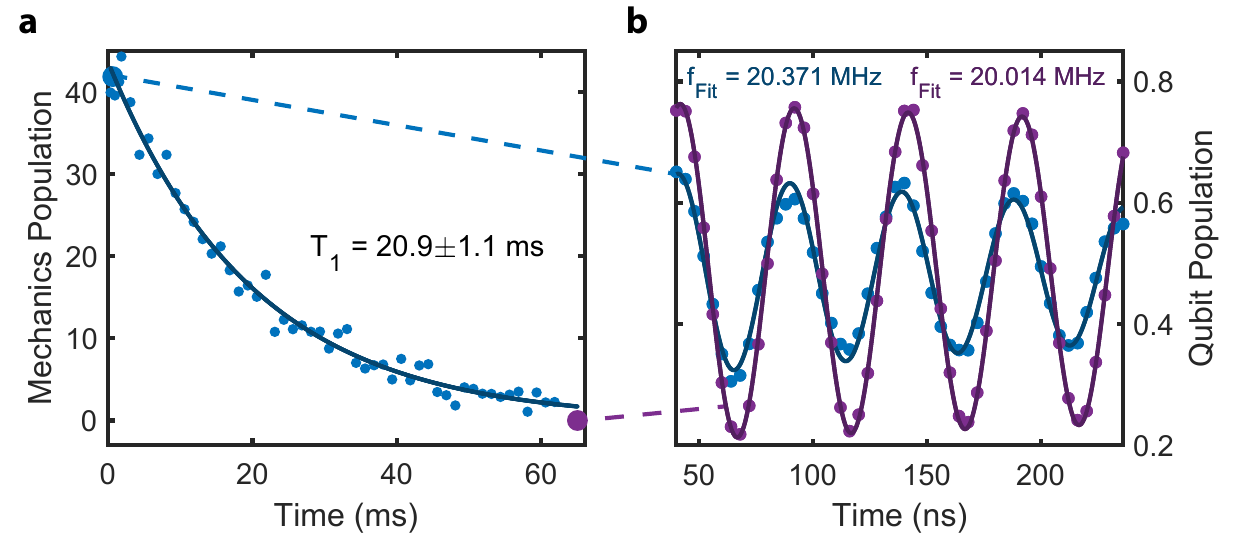}
    \caption{\textbf{Intermediate-phonon-number measurement.} \textbf{a}, Intermediate-phonon-number decay measurement for mechanics A performed at around 40 V applied DC voltage. \textbf{b}, Ramsey fringes for two representative points from the early (blue) and late (purple) stages of the measurement, with measured frequencies of 20.371 and 20.014 MHz, respectively. The solid lines indicate the theory fits used to extract the qubit frequency.}
    \label{fig:Supp_ac_ringdown}
\end{figure}

To measure mechanical lifetime at intermediate phonon numbers, we apply a 1 ms-long tone, resonant with the mechanical oscillator, through the qubit to excite the oscillator. Subsequently, we wait for a varying time interval $t$, followed by a measurement of the qubit frequency. The excitation of the mechanical oscillator induces an ac Stark shift to the qubit transition frequency, described by \cite{blaisCircuitQuantumElectrodynamics2021, vonlupkeParityMeasurementStrong2022}
\begin{equation}
    \delta \omega_\text{q} =  -\frac{2g_\text{em}^2\alpha}{\Delta(\Delta - \alpha)}\bar{n}_\text{phonon} \approx  \frac{2g_\text{em}^2}{\Delta}\bar{n}_\text{phonon}.
    \label{eq:supp_Stark}
\end{equation}
Here, $\bar{n}_\text{phonon}$ is the average phonon number in the oscillator, $\alpha$ is the qubit anharmonicity, $g_\text{em}$ is the electromechanical coupling rate, and $\Delta = \omega_\text{q} - \omega_\text{m}$ represents the detuning between the mechanics and qubit bare frequencies. The approximation holds under the condition $\alpha \gg \Delta$. Consequently, the phonon number in the oscillator can be inferred from measurements of qubit frequency. To extract qubit frequencies, we perform Ramsey measurement, and fit the resulting data to determine the shift in the qubit transition frequency \cite{kristenAmplitudeFrequencySensing2020, schlor2019}. The fitted frequencies relate to the qubit frequency by $\omega_\text{q} = \omega_\text{drive} - \omega_\text{Fit}$, such that $\delta\omega_\text{q} = -\delta \omega_\text{Fit}$. To remove the offset of the drive frequency, we subtract the frequency of the last data point from each data point. An example of this measurement is illustrated in \cref{fig:Supp_ac_ringdown}, which shows the Ramsey measurement results, theoretical fits, and corresponding fitted frequencies for two representative data points.

\subsection{High-phonon-number measurement}
\label{si_high_PhN}

To determine the phonon number in the oscillator, we measure the emitted power from the mechanical oscillator into the waveguide. The relation is given by the expression \cite{bozkurt2023}:
\begin{equation}
    P_\text{out} = n_\text{ph} \frac{\hbar \omega_\text{m} \kappa_\text{e,m}}{2},
\end{equation}
where $P_\text{out}$ is the output power, $n_\text{ph}$ is the phonon number in the oscillator, $\omega_\text{m}$ is the frequency of the mechanical oscillator, and $\kappa_\text{e,m}$ is the external coupling rate of the oscillator to the waveguide. The detected signal consists of two contributions: the desired emission from the mechanical oscillator and the noise from the amplifier. The amplifier noise is characterized and subtracted, allowing the phonon number to be accurately determined. 

In addition to the primary mechanical oscillator (mechanics C), the waveguide to which it is coupled hosts five additional mechanical oscillators. Decay measurements conducted on these oscillators, using the same method as for mechanics C, reveal lifetimes of approximately $20$ ms for multiple devices at high phonon numbers. These measurements exhibit small device-to-device variations, underscoring the reliability of achieving long lifetime in mechanical oscillators. The results of these measurements are summarized in \cref{table:waveguide_T1}.

\begin{table}[h]
\caption{\textbf{Waveguide coupled mechanics. } Lifetime of all mechanics coupled to the waveguide and measured using the high-phonon-number measurement method.}
\label{table:waveguide_T1}
\begin{center}
\begin{tabular}{c c}
mechanics & $T_1$ (ms) \\

\hline
C & $21 \pm 1$ \\
D & $34 \pm 1$ \\
E & $20.1 \pm 0.9$ \\
F & $19.7 \pm 0.6$ \\
G & $22 \pm 1$ \\
H & $16.1 \pm 0.8$ \\
\hline\hline

\end{tabular}
\end{center}
\end{table}

\section{Origin of acoustic loss}
\label{si_acoustic_loss}
Our mechanical devices have been shown to exhibit intrinsic lifetimes as long as $T_\text{1,m}\approx 25 \text{ ms}$, corresponding to an acoustic loss rate of $\Gamma_\text{1,m}/2\pi \approx 6 \text{ Hz}$. Below, we attempt to model the sources of this acoustic dissipation.

\subsection{Acoustic radiation}

We first investigate the loss experienced by the mechanical oscillators due to the radiation of acoustic energy through the clamping points to the substrate \cite{ekinci2005}. The acoustic shields that clamp our oscillators are designed to suppress this effect strongly by having a large band gap ($\approx 2 \text{ GHz}$) roughly centered at the oscillator resonance frequency. In a practical situation, the effectiveness of the phononic shields are generally set by fabrication disorder, which can compromise the performance of these shields, limiting their effectiveness in suppressing acoustic radiation. Experimentally, the acoustic radiation can be investigated through the characterization of devices with different numbers of shields, which provides information about the scaling of radiation loss with the number of shields \cite{maccabe2020}. However, this is a very expensive procedure. 

An alternative method is to rely on numerical simulations of fabrication disorder, which can accurately capture the acoustic radiation in phononic shield structures. Following the same simulation procedure as MacCabe \textit{et al.}, we estimate the acoustic radiation rate for our structure to be $\Gamma_\text{1,m}/2\pi \approx 1 \text{ mHz}$ ($Q > 10^{12}$) (see \cref{si_mech_design}), much smaller than the decay rates measured. Therefore, it is unlikely that acoustic radiation is a significant contributor to the observed loss.

\subsection{Mechanics-TLS interaction}
\label{si_stm}

At our measured bath temperatures of $\sim 70 \text{ mK}$, acoustic loss mechanisms originating from phonon-phonon interactions, such as Landau-Rumer and thermoelastic damping, are expected to be negligible \cite{maccabe2020}. However, defects in the form of TLS are known to be a source of dissipation for acoustic systems at these temperatures \cite{andersson2021,hauer2018,maccabe2020}. In our experiments, we have directly observed signatures of mechanics-TLS interactions in the form of avoided crossings in spectroscopy measurements (see \cref{si_tls_spec}). Furthermore, we have seen that the mechanical lifetime depends on voltage (see \cref{fig:Supp_TLS_loss}a), which is attributed to voltage-dependent TLS frequency shifts. We note that mechanical lifetimes approaching 25 ms can be reliably observed on our devices and represent the lower limit of acoustic dissipation for our system, which constitutes the main focus of our investigations here.

The above indications all suggest that TLS is a potentially significant loss mechanism. We model this contribution to loss by considering the interaction between a mechanical oscillator and TLS defects. For this purpose, we reproduce the results of the standard tunneling model used to describe these defects \cite{muller2019}. Within this model, a TLS is characterized by its asymmetry energy $\epsilon$ and tunneling energy $\Delta_0$, having a transition energy of $E = \sqrt{\epsilon^2+\Delta_0^2}$. The TLS interacts with our oscillator through the strain field associated with mechanical motion. The dynamics of the interaction can be described by a Hamiltonian of the form \cite{behunin2016}
\begin{equation}
    \hat{H}_\text{TLS-m}/\hbar = (g_\text{TLS} \hat{\sigma}_{x,\text{TLS}} +g_{\text{TLS},\ell} \hat{\sigma}_{z,\text{TLS}})  (\hat{b}^\dagger + \hat{b}), \label{eq:sup_TLS_hamiltonian} 
\end{equation}
where $ \hat{\sigma}_{x,\text{TLS}}$ and $ \hat{\sigma}_{z,\text{TLS}}$ are Pauli operators for TLS in the diagonalized energy basis and $g_\text{TLS}$ ($g_{\text{TLS},\ell}$) is the transverse (longitudinal) interaction rate. The interaction rates are given as 
\begin{align}
\begin{split}
       {\hbar g_\mathrm{TLS}} &= \frac{\Delta_0}{E} \tilde{\gamma} \cdot \tilde{\mathbf{S}},\\
        {\hbar g_{\mathrm{TLS},\ell}} &= \frac{\epsilon}{E} \tilde{\gamma} \cdot \tilde{\mathbf{S}},
\end{split}
\end{align}
where $\tilde{\gamma}$ is the TLS deformation potential tensor and $\tilde{\mathbf{S}}$ is the value of the strain tensor at the defect position. As will be detailed in the following sections, both the longitudinal and transverse interactions can give rise to acoustic loss. In order to calculate the TLS-induced acoustic loss, we need quantities such as the density of TLS and the mechanics-TLS interaction rates \cite{behunin2016}, which we estimate here.

First, we look at the TLS spectral density. The nanobeam hosting the mechanical oscillator is expected to host TLS primarily on the silicon surface \cite{woods2019}, as the bulk single-crystal silicon is known to have a low TLS density \cite{kleiman1987}. The surface area of the nanobeam can be calculated to be $12 \,\mu\text{m}^2$. The thickness of the amorphous region on the silicon surface has been shown to be on the order of 3 nm \cite{altoe2022}, giving us a total volume of $V = 0.04 \,\mu\text{m}^3 $ that would host TLS. For a TLS density per unit energy and unit volume of $P_D$, the TLS spectral density is given by $\rho = V P_D$ \cite{ramos2013}. Assuming $P_D\approx 5 \times 10^{44} \,(\text{J.m}^3)^{-1}$ based on the literature \cite{behunin2016}, we expect to have a TLS spectral density of $\rho  \approx 14\,\text{GHz}^{-1}$ interacting with our mechanical oscillator.

Second, we look at the mechanics-TLS interaction rate ($g_\text{TLS}$ and $g_{\text{TLS},\ell}$). The average TLS deformation potential has been characterized to be on the order of 1 eV \cite{behunin2016}. The strain field associated with the mechanical motion can be obtained from numerical modeling using finite element method simulations. These simulations indicate that the average strain field magnitude on the nanobeam surface is $\approx 4\times 10^{-9} \text{ m/m}$. The strain magnitude combined with the deformation potential value suggests that the interaction rate is of the order  $|\tilde{\gamma} \cdot \tilde{\mathbf{S}}|/h \approx 1 \text{ MHz}$.

Armed with estimates of the interaction rates and the TLS density, we will proceed with calculating the expected acoustic loss due to transverse and longitudinal interactions in the following sections, starting with the transverse interaction. We stress that our calculation of the volume that hosts TLS is subject to significant uncertainty because of the strong dependence of the device surface quality on fabrication procedures. Therefore, the following estimates for the TLS-induced acoustic loss are expected to have similarly large uncertainty and possess roughly an order of magnitude level precision.  

\subsection{Resonant TLS loss}
\label{si_tls_loss}
We first focus on loss due to the transverse mechanics-TLS interaction, described in \cref{eq:sup_TLS_hamiltonian}. The transverse interaction gives rise to energy exchange between the mechanical oscillator and the TLS when they are resonant, leading to acoustic loss. This acoustic loss depends on the interaction rate, the frequency detuning between the two parties, and the TLS decay and dephasing rates. For an ensemble of TLS, the total mechanical decay rate can be expressed as \cite{maccabe2020}
\begin{equation}
\label{eq:si_TLS_res}
    \Gamma_\text{1,m} = \sum_\text{TLS} \frac{g_\text{TLS}^2 \Gamma_\text{1,TLS} \tanh{\left(\hbar\omega_\text{TLS}/2k_BT\right)}}{(\omega_\text{TLS}-\omega_\text{m})^2 + (\Gamma_\text{1,TLS}/2+\Gamma_{\phi,\text{TLS}})^2},
\end{equation}
where $\Gamma_\text{1,m} = 1/T_\text{1,m}$ is the decay rate of the mechanics due to transverse interactions with TLS, $\omega_\text{TLS}$ ($\omega_\text{m}$) is the TLS (mechanics) frequency, $g_\text{TLS}$ is the TLS-mechanics transverse interaction rate, $\Gamma_\text{1,TLS}$ is the TLS decay rate, $\Gamma_{\phi,\text{TLS}}$ is the TLS pure dephasing rate and  $k_bT$ is the thermal energy at a temperature of $T$. The decay rate is expressed through a sum that adds up the decay contributions of all the individual TLS interacting with the mechanical oscillator. As can be seen in \cref{eq:si_TLS_res}, the decay is more pronounced for TLS that satisfy the resonance condition with the mechanical oscillator, due to which this loss mechanism due to transverse coupling is referred to as \emph{resonant} TLS loss in the literature \cite{behunin2016,maccabe2020,wollack2021b}.

\begin{figure}[thb]
    \centering
    \includegraphics[width=.6\textwidth]{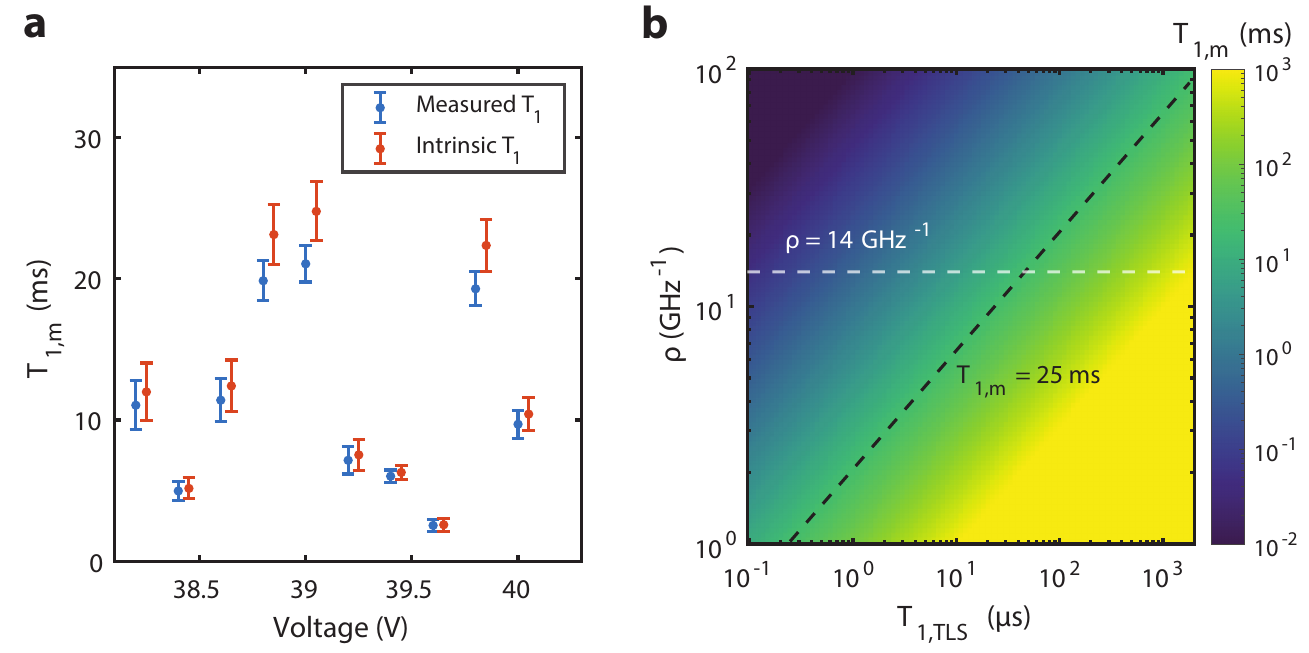}
    \caption{\textbf{Acoustic loss.}  \textbf{a,} Measurement of mechanical lifetime as the voltage is varied in fixed steps. Blue data points represent the measured lifetime of mechanics A for various voltages. Red data points show the intrinsic lifetime obtained by subtracting the inverse Purcell decay from the measured decay rate. \textbf{b,} Mechanical lifetime limit for near-resonant TLS loss as the TLS lifetime and the TLS density is varied. The data is calculated for the best case scenario of $\delta_0 = \delta/2$. The white dashed horizontal line represents the estimated TLS density of $14\text{ GHz}^{-1}$. The black dashed line indicates the TLS lifetime and density combinations that lead to a mechanical lifetime limit of 25 ms, the longest lifetime we have measured. }
    \label{fig:Supp_TLS_loss}
\end{figure}

In a physical system, TLS defects have random positions, orientations, and energies. As a result, several quantities in \cref{eq:si_TLS_res} such as frequency detuning, interaction rate, and the TLS lifetime will be randomly distributed. Therefore, the total decay rate depends on the specific TLS realization that the mechanical oscillator interacts with and is expected to be a random quantity itself. However, it is possible to follow a simplified model to get crude estimates for the mechanical decay rate. In this model, all of the TLS defects are assumed to be interacting with the mechanical resonator at the same coupling rate $\bar{g}_\text{TLS}$ and the frequency separation between the neighboring defects is assumed to be a fixed value of $\delta$. The TLS is also assumed to have a fixed decay rate of $\bar{\Gamma}_\text{1,TLS}$. Further noting that $\tanh{\left(\hbar\omega_\text{TLS}/2k_BT\right)} \approx 1$ in our temperature range where $\hbar\omega_\text{TLS} \gg k_BT$, the expression for the mechanics decay rate can be written as  \cite{odeh2023}
\begin{equation}
    \Gamma_\text{1,m} = \sum_{k=-\infty}^{\infty} \frac{\bar{g}_\text{TLS}^2 \bar{\Gamma}_\text{1,TLS} }{(k\delta+\delta_0)^2 + (\bar{\Gamma}_\text{1,TLS}/2)^2},
    \label{eq:si_purcell}
\end{equation}
where $\delta_0 \in \{0,\delta/2\}$ is the frequency separation between the mechanics and the nearest TLS and the dephasing of the TLS has been neglected. We use an average interaction rate of $\bar{g}_\text{TLS}/2\pi = 1 \text{ MHz}$ (where the average tunneling energy ratio is $\Delta_0/E\approx 1$ \cite{ramos2013}). For a best case TLS distribution where the closest TLS is away by $\delta_0 = \delta/2$, we can calculate the mechanical lifetime due to the resonant TLS for a given set of TLS densities and TLS lifetimes (see \cref{fig:Supp_TLS_loss}b). For our device, we estimate $\delta/2\pi = \rho^{-1} = 70 \text{ MHz}$ which is set by the TLS density of $14 \text{ GHz}^{-1}$.  TLS lifetimes have been measured to be $1/\bar{\Gamma}_\text{1,TLS} = T_\text{1,TLS} \approx 0.1-1 \,\mu\text{s}$ in silicon \cite{lisenfeld2016,shalibo2010}. Using these numbers, we would expect to have a mechanical lifetime of $1/\Gamma_\text{1,m} = T_\text{1,m}\approx 50-500 \,\mu\text{s}$, 2 orders of magnitude smaller than the observed lifetime of 25 ms.

However, this apparent contradiction can be resolved by noting that TLS lifetimes can be modified through phononic engineering. Specifically, since the TLS relaxation occurs mainly through interaction with phonons, the TLS decay rate is related to the phononic density of states (DOS), which is a function of the device geometry \cite{behunin2016}. Comparable devices that utilize phononic shields have shown enhanced TLS lifetimes (average lifetime of $\approx 400 \,\mu\text{s}$) through suppression of the phononic DOS inside the bandgap \cite{chen2024}. The combination of the previously measured TLS lifetimes and our simulations indicating a $2 \text{ GHz}$ wide phononic bandgap (see \cref{si_mech_design}) can be used to estimate the acoustic loss due to TLS inside the bandgap to be $\Gamma_\text{1,m} \approx 2\pi\times 0.8 \text{ Hz}$. Notably, this number is smaller than the observed decay rate of $\Gamma_\text{1,m} \approx 2\pi\times 6 \text{ Hz}$. Importantly, resonant loss from these TLS is expected to be saturable \cite{behunin2016}. The absence of saturable loss in our measurements (see \cref{fig:lifetime}) further corroborates that the resonant loss due to TLS inside the bandgap is minimal.

The TLS outside the bandgap, on the other hand, is expected to have shorter lifetimes and have been measured to possess and average lifetime of $\approx 4 \,\mu\text{s}$. Similarly using this average lifetime, the loss due TLS outside the bandgap can be estimated to be $\Gamma_\text{1,m} \approx 2\pi\times 1 \text{ Hz}$, a number comparable to the TLS loss inside bandgap.

\subsection{Relaxation damping}

We next explore the loss due to the longitudinal interaction between mechanics and TLS. The longitudinal interaction in  \cref{eq:sup_TLS_hamiltonian}, manifests itself as a modulation of the TLS energy due to the acoustic displacement, moving it out of thermal equilibrium. The TLS relaxes towards thermal equilibrium, giving rise to acoustic loss at nonzero temperatures. The acoustic loss due to this process is called relaxation damping, whose magnitude can be expressed as \cite{behunin2016}
\begin{equation}
    \Gamma_\text{1,m} = \sum_\text{TLS} \frac{2g_{\text{TLS,}\ell}^2}{\omega_\text{m}} \frac{\hbar\Gamma_\text{1,TLS}}{k_BT} \text{sech}^2{\left(\hbar\omega_\text{TLS}/2k_BT\right)}.
    \label{eq:si_relax}
\end{equation} 
Due to the $\text{sech}^2{\left(\hbar\omega_\text{TLS}/2k_BT\right)}$ term, the loss is mainly dominated by thermally active TLS, having energies of  $\hbar\omega_\text{TLS} \lesssim k_B T$. At our measured bath temperature of $T\sim70 \text{ mK}$, this corresponds to $\omega_\text{TLS}/2\pi \lesssim 1.4 \text{ GHz}$.

Similar to the procedure we have followed for resonant loss, we can numerically estimate the loss by assuming that the TLS have a uniform frequency separation of $\delta /2\pi = 70 \text{ MHz}$ and interact with the oscillator with the average longitudinal coupling rate $\bar{g}_{\text{TLS,}\ell}/2\pi \approx 0.6 \text{ MHz}$ (where the average asymmetry energy ratio is $\epsilon/E\approx 0.6$ \cite{ramos2013}). The remaining parameter needed for calculation is the TLS decay rate. The decay rates for TLS depends on the phononic DOS at the TLS frequency as discussed above. Specifically, at the frequencies of the thermally active TLS we expect our phononic DOS to deviate from the 3D case, as the dimensions of our beam ($\sim 0.2 \,\mu\text{m} $ thickness and $\sim 0.5 \,\mu\text{m} $ width) are much smaller than the acoustic wavelength of $\gtrsim 5 \,\mu\text{m}$. Hence, the decay rates of the TLS can be attributed to a phononic system with dimensionality between 1D and 2D \cite{maccabe2020}, which lends itself to analytical formulas for idealized systems \cite{behunin2016}.  Using these decay rates, we can calculate the relaxation damping rate to be $\Gamma_\text{1,m} \approx 2\pi \times 0.7 \text{ Hz}$. This decay rate is comparable to the estimates for resonant loss. 

In the preceding analysis, we have used identical densities for high-frequency TLS that cause near-resonant loss and low-frequency thermally active TLS that lead to relaxation damping. However, it should be noted that there have been observations of thermally active TLS densities being larger due to an abundance of surface spins \cite{degraaf2018}. This suggests that our calculations may underestimate the loss due to relaxation damping.

The total sum of the estimated loss rates from various interactions with TLS amounts to approximately $\Gamma_\text{1,m} \approx 2\pi \times 2.5 \text{ Hz}$. Although this number is subject to a large uncertainty (because of the order-of-magnitude nature of the estimates), we note that it is in reasonable agreement with the observed loss $2\pi \times 6 \text{ Hz}$.  In order to disentangle the contribution of each loss mechanism, future studies that systematically investigate the temperature and power dependence of the mechanical quality factor are needed \cite{wallucks2020}.

\section{Frequency noise}
\label{si_freq_noise}
\subsection{Dephasing theory}
\label{si_dephasing}
For the Gaussian theory of dephasing, we reproduce previous analyses \cite{ithier2005}. The frequency noise is commonly investigated in physical systems via measurements of free induction and echo. The  free induction signal ($f_{z,R}$) is related to the frequency noise $\delta\omega$ through the random phase accumulated \begin{equation}
    \Delta\phi = \int_0^t dt' \delta\omega(t'),
\end{equation} where $f_{z,R} = \langle\exp(i\Delta\phi)\rangle$. For a Ramsey free induction measurement, the signal $f_{z,R}$ corresponds to the envelope of the oscillations.  Similarly, the echo signal $f_{z,E}$ is related to the frequency noise by an echo phase, that includes the refocusing effect of the echo experiment \begin{equation}
    \Delta\phi_E = \int_0^{t/2} dt' \delta\omega(t') - \int_{t/2}^{t} dt' \delta\omega(t'), 
\end{equation}
where $f_{z,E} = \langle\exp(i\Delta\phi_E)\rangle$.

In the presence of Gaussian frequency noise with power spectral density $S_{\delta\omega}(\omega)$, the free induction ($f_{z,R}$) and echo ($f_{z,E}$) signals can be expressed as
\begin{align}
    \ln{(f_{z,R}(t))} &= x_R(t) = -\frac{t^2}{2} \int d\omega S_{\delta\omega}(\omega) \text{sinc}^2 \frac{\omega t}{2}, \\
    \ln{(f_{z,E}(t))} &= x_E(t) = -\frac{t^2}{2} \int d\omega S_{\delta\omega}(\omega) \text{sin}^2\frac{\omega t}{4}\text{sinc}^2 \frac{\omega t}{2}.
\end{align}

\begin{figure}[ht]
    \centering
    \includegraphics[width=.6\textwidth]{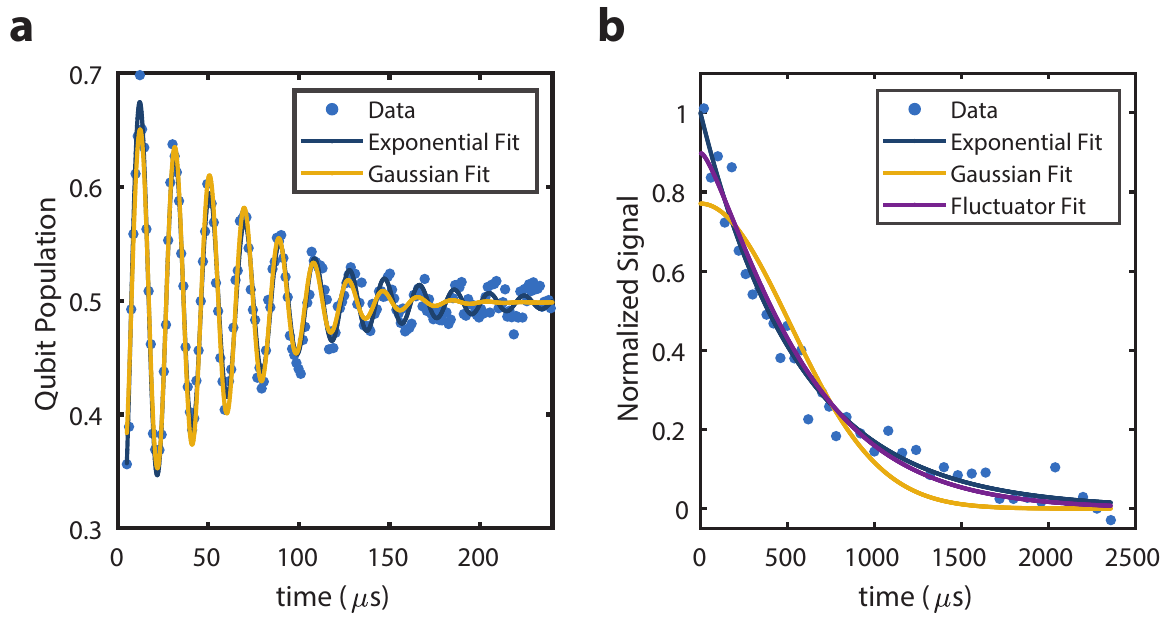}
    \caption{\textbf{Ramsey and echo fits.} \textbf{a,} Exponential and Gaussian fits to the Ramsey signal showing that the exponential fit better captures the coherence at long times. \textbf{b,} Different theory fits to the Hahn echo signal, with the exponential fit from the main text displayed with a Gaussian fit and a fluctuator model fit with $\gamma_\text{max} = 10^5 \text{ s}^{-1}$. }
    \label{fig:sup_ramsey_echo}
\end{figure}

For white Gaussian noise, these results lead to an exponential decay with $x_R \approx x_E \approx t S_{\delta\omega}(0)/2 $. Therefore, the echo process is expected to be inefficient in the case of white noise. Noise that is concentrated at low frequencies is needed for the efficiency of the echo procedure. For example, Gaussian $1/f$ noise with $S_{\delta\omega}(\omega) = A /|\omega|$ gives rise to free induction and echo signals of the form
\begin{align}
    x_R(t) &\approx -t^2 A \ln{\left(\frac{1}{\omega_\text{ir}t}\right)}, \\
    x_E(t) & = -t^2 A \ln{2},
\end{align}
where $\omega_\text{ir}$ is the infrared cutoff frequency, generally identified with the reciprocal of the averaging time. Therefore, for Gaussian $1/f$ noise, both Ramsey and echo signals have Gaussian functional forms and the echo lifetime exceeds the Ramsey lifetime by a factor of 
\begin{equation}
    \frac{T_2^E}{T_2^*} \approx \sqrt{\frac{\ln{\left(\frac{1}{\omega_\text{ir}T_2^*}\right)}}{\ln2}}.
\end{equation}
Despite observing $1/f$ type frequency noise (see \cref{si_psd}) and seeing strong improvements with the echo procedure, the functional forms of our measured Ramsey and echo signals are better described by exponential functions and are not Gaussian (see \cref{fig:sup_ramsey_echo}a,b), inconsistent with the predictions for Gaussian $1/f$ noise.

These observations are not too surprising, as non-Gaussian noise, where the power spectral density is not sufficient to describe the echo and Ramsey signals, is common in physical systems \cite{szankowski2017a,galperin2006a}.  For example, the ``telegrapher'' frequency noise model, where the frequency switches between two stable states $\pm \nu$ with a switching rate $\gamma$, is Gaussian only under the specific limit of $\nu\ll\gamma$ \cite{bergli2009}. In physical systems, thermal fluctuators, which are thermally active two-level-systems ($E\lesssim k_bT$) can give rise to telegrapher noise. In fact, noise from a collection of thermal fluctuators can give rise to 1/f noise \cite{bergli2009}. However, for the description of the free induction and echo signals, a distribution of the telegrapher noise processes is needed.

To this end, we follow an analysis that models the noise as an ensemble of telegrapher noise processes caused by thermal fluctuators, with amplitudes ($\nu$) and rates ($\gamma$) having a distribution of \begin{equation}
    P(\nu,\gamma) = \frac{\xi}{\gamma\nu^2},
\end{equation}
motivated by the underlying physics of two-level systems \cite{galperin2004,ithier2005}. The distribution is defined in the domain $[\nu_\text{min}, \infty] \times [\gamma_\text{min},\gamma_\text{max}] $ and $\xi$ is a normalization parameter. For thermal fluctuators, the switching rates are set by the decay rates of the thermally active TLSs, which are set by interacting with acoustic phonons \cite{faoro2015}. $\gamma_\text{min}$ is set by the switching rate of the slowest fluctuator in the ensemble (observed to be $\lesssim 1 \text{ mHz}$ \cite{faoro2015}) or the reciprocal of the observation duration (smallest amongst the two). $\gamma_\text{max}$ is the switching rate of the fastest fluctuator in the ensemble. At our bath temperatures of $\sim 70 \text{ mK}$, the thermally active TLS are expected to have frequencies $\omega/2\pi \lesssim 1.4 \text{ GHz}$. The fastest fluctuators are expected to have a rate of $\sim 10^5 \text{ s}^{-1}$ based on previous observations \cite{shalibo2010}. First principle calculations for the TLS decay rate in the presence of 1D or 2D acoustic baths of our system gives similar rates for the fastest fluctuator of $\sim 10^5 \text{ s}^{-1}$ \cite{behunin2016}.

\begin{figure}[ht]
    \centering
    \includegraphics[width=.35\textwidth]{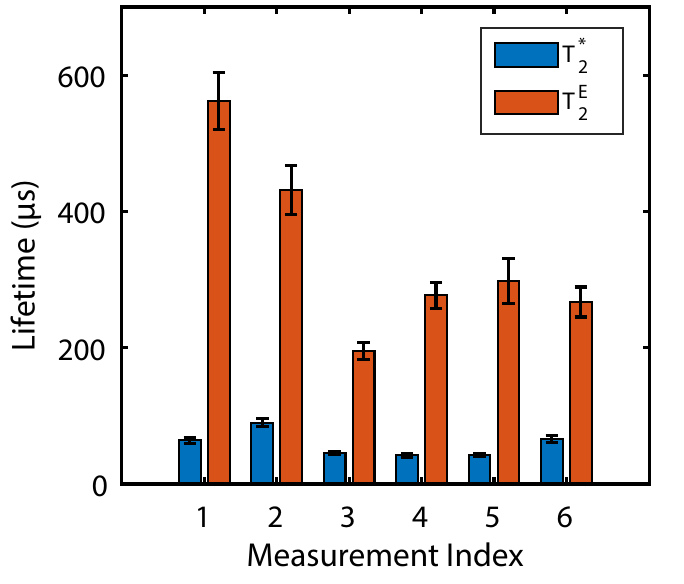}
    \caption{\textbf{Ramsey and Hahn echo lifetimes.} Measurements of Ramsey and Hahn echo lifetimes for mechanical oscillators A and B at different voltage operation points (labeled with a measurement index between 1-6), demonstrating the consistent efficiency of the echo procedure in removing frequency noise. The error bars indicate the $2\sigma$ confidence interval. }
    \label{fig:sup_echo_enhancement}
\end{figure}

This distribution gives rise to noise with a $1/f$ power spectral density.  The corresponding Ramsey and echo signals are given as 
\begin{align}
    x_R(t) &\approx -t \xi \ln{\left(\gamma_\text{max}/\gamma_\text{min}\right)}, \\
    x_E(t) & \approx -t \xi \ln(\gamma_\text{max}t), \label{eq:echo_func}
\end{align}
where the expression for the echo signal is valid in the range $t>\gamma_\text{max}^{-1}$. The Ramsey signal has the exponential form noted in our experiments. The expression for the echo is valid due to the dynamics of the most rapid fluctuator in the system being faster than the echo timescale  ($t\sim T_2^E \approx 500 \text{ us} >\gamma_\text{max}^{-1} \approx 10  \text{ us}  $). We observe that a fit to our data using the echo function in \cref{eq:echo_func} with $\gamma_\text{max} = 10^5 $ shows good agreement, as seen in \cref{fig:sup_ramsey_echo}b. It is important to note that the echo and Ramsey signals depend on the specific set of fluctuators that a device interacts with, which is in line with our observations of Ramsey and echo times showing variance with device and voltage (see \cref{fig:sup_echo_enhancement}). The voltage dependence can be potentially attributed to the TLS frequencies getting Stark shifted due to the electrostatic bias field, altering the fluctuator ensemble that the device interacts with \cite{lisenfeld2016,lisenfeld2019}. 

The echo efficiency predicted by this model can be expressed as 
\begin{equation}
    \frac{T_2^E}{T_2^*} \approx \frac{\ln{\left(\gamma_\text{max}/\gamma_\text{min}\right)}}{\ln(\gamma_\text{max}T_2^E)}.
\end{equation}
The thermal fluctuators can exhibit a wide range of switching rates, as discussed above, leading to $\ln{(\gamma_\text{max}/\gamma_\text{min})}\approx 20$ \cite{faoro2015}. This results in an estimated echo efficiency of $\approx$ 5 for our devices. The estimate is in line with our average improvement of $\approx 6$ (see \cref{fig:sup_echo_enhancement}). Therefore, we see that the analysis we have followed can explain the observed power spectral density, functional forms of the Ramsey and echo signals, the improved lifetimes in the presence of dynamical decoupling and voltage dependence of the lifetimes. We note that the probability distribution of the telegrapher noise processes, despite being physically motivated, is phenomenological and a more systematic analysis would have to depart from TLS-TLS interactions as a starting point \cite{faoro2015}.

\subsection{Power spectral density}
\label{si_psd}

Mechanical frequency noise can be investigated through repeated measurements of the oscillation frequency, where the noise manifests itself as frequency fluctuations. A method to measure the frequency of the mechanical oscillator is to rely on the fringes in Ramsey measurements. Specifically, the fringe frequency is set by the detuning between the drive tone and the mechanical frequency, providing a way to infer the mechanical frequency for a known drive frequency. Utilizing this method, we investigate the frequency noise through long measurements.

\begin{figure}[ht]
    \centering
    \includegraphics[width=.8\textwidth]{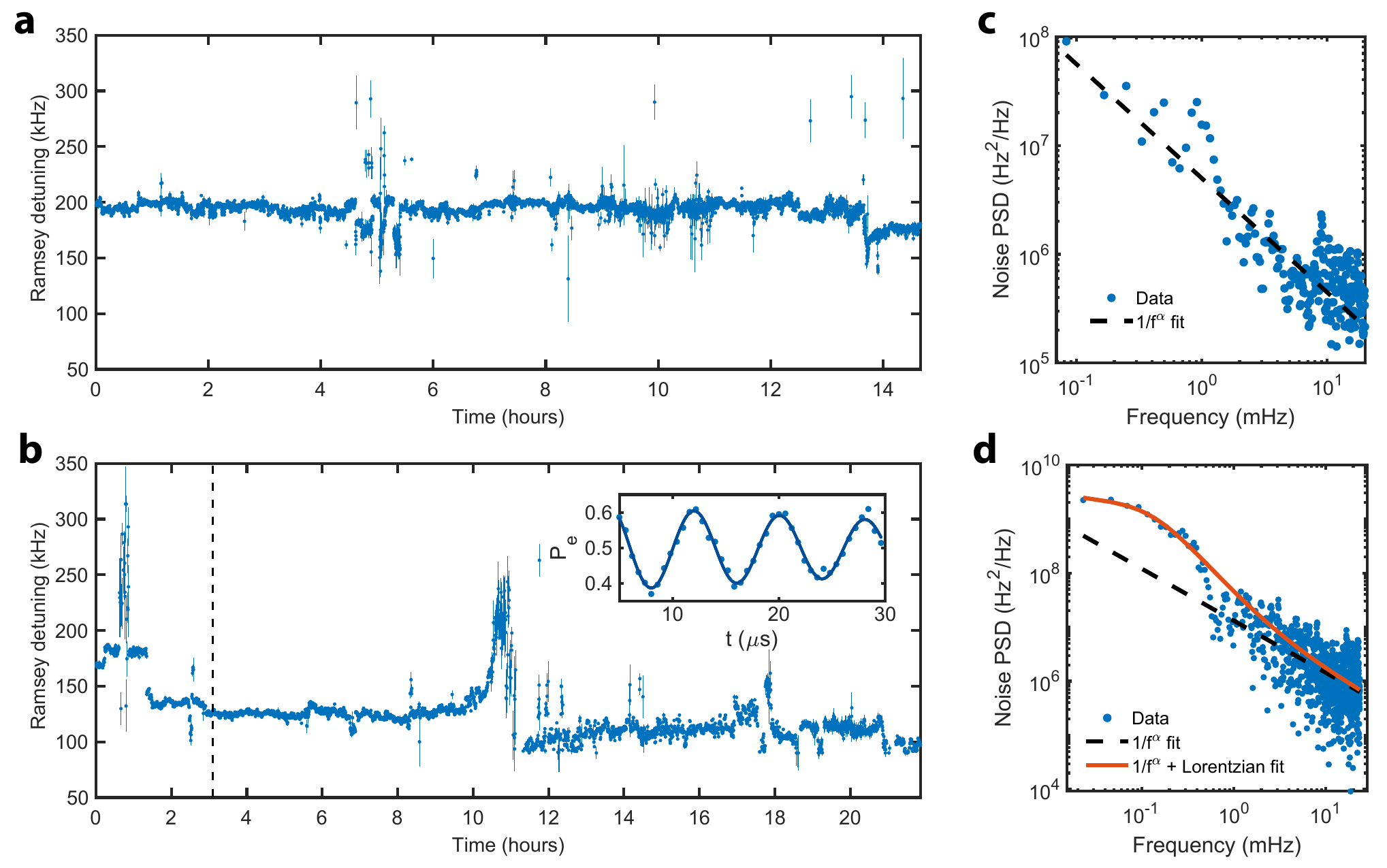}
    \caption{\textbf{Frequency fluctuations.} \textbf{a,} (\textbf{b,}) Ramsey frequency detuning of mechanics A at 50 (40) Volt applied DC voltage. As shown with an example as the inset, the Ramsey frequency is calculated using fits to the Ramsey measurement of the mechanics via the qubit. \textbf{c,} The frequency noise PSD of data in panel a, revealing $1/f^\alpha$ type noise with $\alpha = 1.0\pm0.1$. \textbf{d,}  PSD of the data in panel b, demonstrating $1/f^\alpha$ type noise (with $\alpha = 0.96\pm0.08$) at large frequencies and added Lorentzian noise at low frequencies attributed to a dominant TLS.}
    \label{fig:sup_PSD}
\end{figure}

Time-dependent dynamics of oscillation frequency is presented in \cref{fig:sup_PSD}a and b, for different voltage biases of 50 V and 40 V. The 50 V point corresponds to the data presented in the main text ( \cref{fig:coherence}a,c and d) and the 40 V point is chosen to be in the near spectral vicinity of a TLS. Through analysis of the time traces, the power spectral density of the frequency noise can be obtained. The measurements are conducted approximately every 20 seconds, which sets an upper bound on the frequency of the PSD that can be captured without aliasing, known as Nyquist frequency. The power spectral density (PSD) of the frequency at 50 V, as presented in \cref{fig:sup_PSD}c, reveals a $1/f^\alpha$ type noise, a characteristic feature in a variety of solid-state systems \cite{paladinoNoiseImplicationsSolidstate2014}. This type of frequency noise is quite often attributed to a bath of two-level systems, which we have utilized in modeling our dephasing (see \cref{si_dephasing}). Interestingly, for the PSD at 40 V (\cref{fig:sup_PSD}d), we see deviations from $1/f$ noise. The PSD is in fact better explained by a Lorentzian added to the $1/f^\alpha$ noise floor. The Lorentzian corresponds to a telegrapher noise process, whose PSD is given the functional form
\begin{equation}
    S(\omega) \propto \frac{2\gamma_1}{\gamma_1^2 + \omega^2},
\end{equation}
where $\gamma_1$ is the switching rate. From the fit, we extract a switching rate of $\gamma_1 = 0.1 \pm 0.002$ mHz.

Importantly, individual TLS are known to cause telegrapher-type frequency noise, where the switching rate of the process corresponds to the relaxation rate of the TLS \cite{faoro2015}. Normally, the $1/f$ noise is obtained from the superposition of numerous Lorentzian processes. However, there can be instances where a single strongly coupled TLS dominates the noise \cite{schlor2019}, leading to a pronounced Lorentzian feature. The observed PSD can be attributed to a similar process. We finally observe that the measured frequency noise changes significantly based on the applied voltage. This can be explained by the TLS landscape changing due to the TLS frequency shifts associated with electrostatic fields (see \cref{si_tls_spec}). This further supports the hypothesis that the mechanical frequency noise is caused by TLS.

\section{TLS Spectroscopy}
\label{si_tls_spec}

\begin{figure}[ht]
    \centering
    \includegraphics[width=\textwidth]{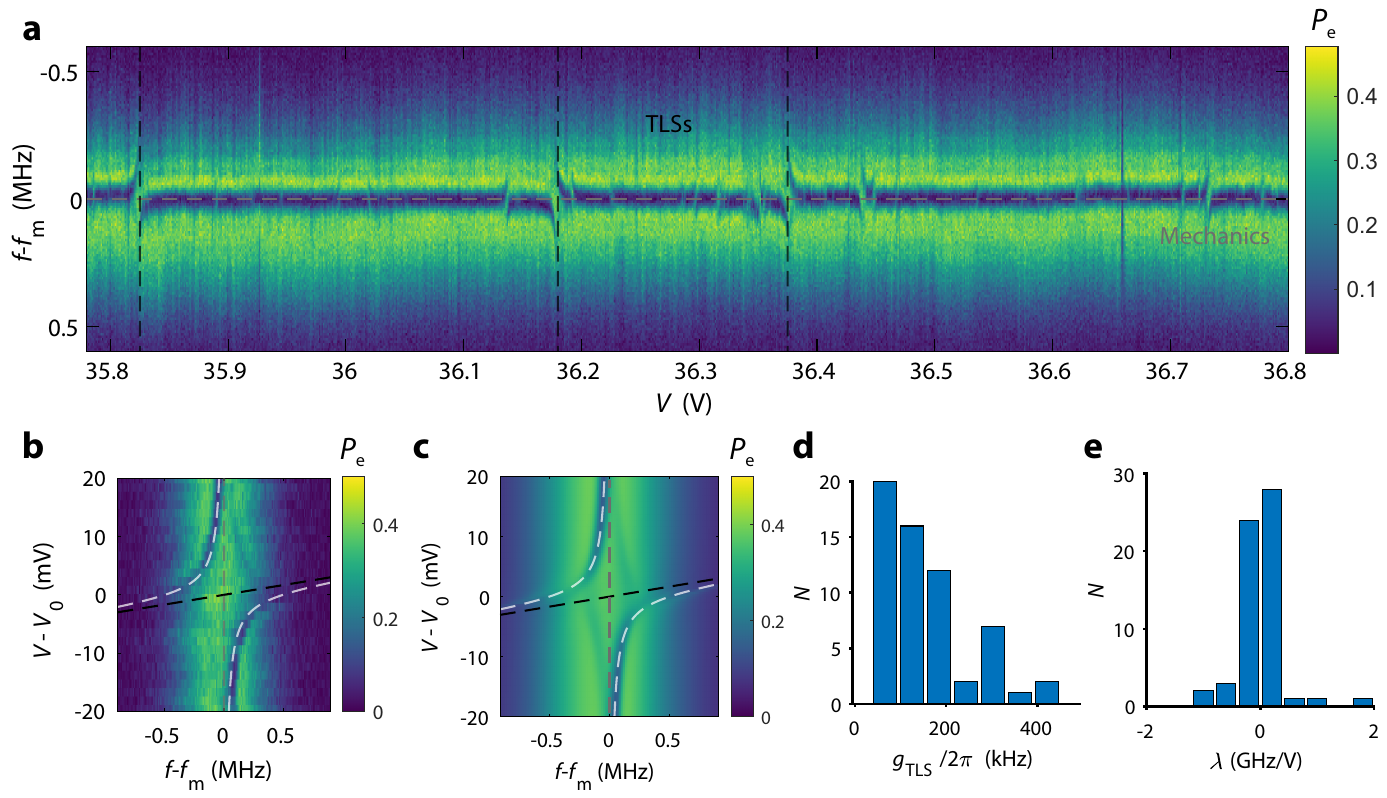}
    \caption{\textbf{TLS spectroscopy.} \textbf{a,} Spectroscopy trace of the qubit-mechanics system as the voltage is varied, illustrating the coupling between the mechanics and individual TLSs through multiple avoided crossings. The black dashed lines highlight a selection of prominent individual TLSs. \textbf{b,} A magnified view of a specific mechanics-TLS avoided crossing. The white dashed lines represent a theory fit to the avoided crossing. The black dashed line is the bare TLS frequency. \textbf{c,} Avoided TLS-mechanics mode crossing obtained by solving the full master equation (with parameters $\lambda \approx 0.3\text{ GHz/V}$ and $ g_\text{TLS}/2\pi \approx 0.5 \text{ MHz}$), confirms the features seen in experimental data. \textbf{d,e} Histograms of the TLS-mechanics interaction rate and TLS tuning rate obtained from avoided crossing fits. $N$ is the number of TLS within a bin.}
    \label{fig:TLS_sweep}
\end{figure}

During our TLS spectroscopy measurements, we have seen numerous individual TLS interact with our mechanical oscillator, as shown in \cref{fig:TLS_sweep}a. We can model the result of spectroscopy measurements through the Hamiltonian
\begin{equation}
    \hat{H}/\hbar = \frac{\omega_\text{q}}{2}\hat{\sigma}_\text{z}  + \frac{\omega_\text{TLS}}{2}\hat{\sigma}^\text{TLS}_\text{z} + \omega_\text{m}\hat{b}^{\dagger}\hat{b} + g_\text{em}\left(\hat{\sigma}_+\hat{b}+\hat{\sigma}_-\hat{b}^{\dagger}\right) + g_\text{TLS}\left(\hat{\sigma}^\text{TLS}_+\hat{b}+\hat{\sigma}^\text{TLS}_-\hat{b}^{\dagger}\right) + \frac{\Omega
    _\text{q}}{2}\left(\hat{\sigma}_++\hat{\sigma}_-\right),
\end{equation}
where $\omega_\text{TLS}$ is the TLS frequency, $\hat{\sigma}^\text{TLS}_-$ is ($\hat{\sigma}^\text{TLS}_+$) the TLS lowering (raising) operator, $g_\text{TLS}$ is the TLS-mechanical oscillator transverse coupling rate and $\Omega_\text{q}$ is the qubit Rabi drive strength. Notably, the TLS frequency in the near vicinity of the mechanical oscillator frequency can be expressed as 
\begin{equation}
    \omega_\text{TLS} = \omega_\text{m} + 2\pi\times\lambda (V-V_0),
\end{equation}
where  $\lambda$ is the TLS tuning rate per volt, $V$ is the applied voltage and $V_0$ is the voltage bias value at which the TLS is in resonance with the mechanical oscillator. The tuning rate $\lambda$ for a TLS can be further expressed as \cite{ramos2013} 
\begin{equation}
\label{eq:tuning_rate}
    h\lambda = \frac{\epsilon}{E} \mathbf{p} \cdot \mathbf{E},
\end{equation}
where $\mathbf{p}$ is the TLS dipole moment and $\mathbf{E}$ is the electric field at the position of the TLS with 1 V applied to the electrodes. We see that we can faithfully reproduce the measurement results through master equation simulations in the steady-state (compare the measurement result in \cref{fig:TLS_sweep}b to the simulation results in \cref{fig:TLS_sweep}c). We note that in the measurements, the qubit corresponds to the broad green feature and the mechanical oscillator manifests itself as dark lines, which physically corresponds to a dip in the spectroscopy response at the mechanical oscillator frequency.

As demonstrated above, the spectroscopy traces can be utilized to extract TLS parameters such as the interaction rate and the tuning rate. Ideally, this procedure can be applied to all of the observed TLS to obtain statistics. However, we encounter experimental challenges in fully resolving the avoided crossings. Mainly, the splitting of the hybridized modes ($2g_\text{TLS}$) can place them outside of the qubit linewidth (the qubit frequency is kept fixed in the experiment), or the signal associated with hybridized modes can be too weak to observe due to decoherence induced by TLS. Therefore, we cannot reliably observe the entirety of the avoided crossing for all of the individual TLS, instead we only see the tail ends of the avoided crossing for some of them. The change in the mechanical frequency due to the TLS for the tail end ($\Delta_\text{TLS}\gg g_\text{TLS}$) can be expressed as $-g_\text{TLS}^2/\Delta_\text{TLS}$ , where $\Delta_\text{TLS} = \omega_\text{TLS}-\omega_\text{m} = 2\pi\times\lambda (V-V_0)$. Since both the interaction rate $g_\text{TLS}$ and the tuning rate $\lambda$ are unknown, we can extract both of these parameters only under the assumption of finding the lower bounds on both of them. Nonetheless, we will follow this procedure for the TLS where only the tail ends are visible, and obtain statistics concerning the lower bounds, which still include valuable information. For the following discussions concerning the statistics of tuning rates and coupling rates, the quantities analyzed will be these lower bounds. 

Exploring a tuning range of 5 V, between 35V and 40 V, we observe avoided crossings with 60 TLS, 56 of which have a coupling rate above $2\pi\times 50 \text{ kHz}$, with the distribution of the coupling rates demonstrated in \cref{fig:TLS_sweep}d. Based on the observed interaction rates, we can infer that TLS exhibit deformation potential values extending to 0.5 eV (see \cref{si_stm} for acoustic strain calculations), consistent with results from literature \cite{behunin2016,grabovskij2012}.  The tuning rate distribution is shown in \cref{fig:TLS_sweep}e, where 31 of them have positive tuning rates and 29 of them have negative tuning rates, in line with the expected equal probability based on random dipole orientations and TLS density. The number of observed TLS combined with the average tuning rate magnitude ($0.22 \text{ GHz/V}$) can be used to obtain an upper bound of 13.8 $\text{GHz}^{-1}$ on the TLS density ($\rho = P_D V$).  This upper bound is compatible with the density value of 14 $\text{GHz}^{-1}$ we have obtained from first principles in \cref{si_stm} and utilized for our calculations of the decay rate.

\twocolumngrid
\bibliography{transmon_mech_bibliography.bib}

\end{document}